\documentclass[10pt, onecolumn]{IEEEtran}

\IEEEoverridecommandlockouts
\usepackage{xspace,exscale,relsize}
\usepackage{fancybox,shadow}
\usepackage{graphicx}
\usepackage{color}
\usepackage{amsfonts}
\usepackage{amssymb}
\usepackage{url}

\usepackage[noadjust]{cite}

\usepackage{amsmath}
	\makeatletter
	\let\over=\@@over \let\overwithdelims=\@@overwithdelims
	\let\atop=\@@atop \let\atopwithdelims=\@@atopwithdelims
  	\let\above=\@@above \let\abovewithdelims=\@@abovewithdelims
  	\makeatother
\usepackage{fixltx2e}

\usepackage{rotating}

\usepackage{ifpdf}

\usepackage{subfigure}
\usepackage{psfrag}

\newcommand{\matc}{\ensuremath{\mathcal{C}}}
\newcommand{\mate}{\ensuremath{\mathcal{E}}}
\newcommand{\matx}{\ensuremath{\mathcal{X}}}

\newcommand{\mata}{\ensuremath{\mathcal{A}}}

\newcommand{\maty}{\ensuremath{\mathcal{Y}}}
\newcommand{\matn}{\ensuremath{\mathcal{N}}}

\newcommand{\mcost}{\ensuremath{\mathtt{c}}}

\newcommand{\sff}{{\sf f}}
\newcommand{\sfg}{{\sf g}}

\newcommand{\mreals}{\ensuremath{\mathbb{R}}}

\ifx\eqref\undefined
	\newcommand{\eqref}[1]{~(\ref{#1})}
\fi
\ifx\mod\undefined
	\def\mod{\mathop{\rm mod}}
\fi

\newcommand{\vect}[1]{{\bf #1}}

\newcommand{\norm}[1]{{\left\Vert #1 \right\Vert}}

\def\exp{\mathop{\rm exp}}

\def\tr{\mathop{\rm tr}}

\DeclareMathOperator\EE{\rm\mathbb{E}}
\DeclareMathOperator\Var{\rm Var}

\def\PP{\mathbb{P}}

\def\eqdef{\stackrel{\triangle}{=}}

\def\bSigma{\mathbf{\Sigma}}
\def\bV{\mathbf{V}}
\def\bIn{\mathbf{I}_n}

\def\unifto{\mathop{{\mskip 3mu plus 2mu minus 1mu%
	\setbox0=\hbox{$\mathchar"3221$}%
	\raise.6ex\copy0\kern-\wd0%
	\lower0.5ex\hbox{$\mathchar"3221$}}\mskip 3mu plus 2mu minus 1mu}}

\ifx\lesssim\undefined
\def\simleq{{{\mskip 3mu plus 2mu minus 1mu%
	\setbox0=\hbox{$\mathchar"013C$}%
	\raise.2ex\copy0\kern-\wd0%
	\lower0.9ex\hbox{$\mathchar"0218$}}\mskip 3mu plus 2mu minus 1mu}}
\else
\def\simleq{\lesssim}
\fi

\ifx\gtrsim\undefined
\def\simgeq{{{\mskip 3mu plus 2mu minus 1mu%
	\setbox0=\hbox{$\mathchar"013E$}%
	\raise.2ex\copy0\kern-\wd0%
	\lower0.9ex\hbox{$\mathchar"0218$}}\mskip 3mu plus 2mu minus 1mu}}
\else
\def\simgeq{\gtrsim}
\fi

\newif\ifincludenorms
\includenormstrue

\newtheorem{remark}{Remark}
\newtheorem{theorem}{Theorem}
\newtheorem{lemma}[theorem]{Lemma}
\newtheorem{corollary}[theorem]{Corollary}
\newtheorem{definition}{Definition}
\newtheorem{proposition}[theorem]{Proposition}

\begin{document}

\title{Empirical distribution of good channel codes with non-vanishing error probability}
\author{Yury~Polyanskiy 
        and~Sergio~Verd\'u 
\thanks{Y. Polyanskiy is with the Department of Electrical Engineering and Computer
Science, MIT, Cambridge, MA, 02139 USA, e-mail: {\ttfamily yp@mit.edu}. S.~Verd\'u is with the Department
	of Electrical Engineering, Princeton University, Princeton, NJ, 08544 USA.
	\mbox{e-mail:~{\ttfamily verdu@princeton.edu}.}}%
\thanks{
The work was supported in part by the National Science Foundation (NSF) under Grant CCF-1016625
and by the Center for Science of Information (CSoI), an NSF Science and Technology Center, under
Grant CCF-0939370. Parts of this work were presented at the 49th and 50th Allerton Conferences on
Communication, Control, and Computing, 2011-2012.
}
}
%

\maketitle

\begin{abstract}
This paper studies several properties of channel codes that approach the fundamental limits of a given
(discrete or Gaussian) memoryless channel with a non-vanishing probability of error. The output distribution induced by an
$\epsilon$-capacity-achieving code is shown to be 
close in a strong sense to the capacity achieving output distribution. 
Relying on the concentration of measure (isoperimetry) property enjoyed by the latter, it is shown
that regular (Lipschitz) functions of channel outputs can be precisely estimated and turn out to be
essentially non-random and independent of the actual code. 
It is also shown that 
the
output distribution of a good code and the capacity achieving one cannot
be distinguished with exponential reliability.
The random process produced at the output of the channel is shown to satisfy the
asymptotic equipartition property. 
\ifincludenorms
Using related methods it is shown that quadratic forms and sums of $q$-th powers when
evaluated at codewords of good AWGN codes approach the values obtained from a randomly generated
Gaussian codeword.
\fi
\end{abstract}


%
\IEEEpeerreviewmaketitle

\section{Introduction}

A reliable channel codebook (or code, for the purposes of this paper) is a collection of
codewords of fixed duration distinguishable with small probability of 
error when observed through a noisy channel. Such a code is optimal
if it possesses the maximal cardinality among all codebooks of equal duration and
probability of error.
In this paper, we characterize several properties of optimal and
close-to-optimal channel codes indirectly, i.e. without identifying the best code explicitly. This
characterization provides theoretical insight and ultimately 
may facilitate the search for new good code families by providing a necessary condition they must satisfy.

Shannon~\cite{CS48} was the first to recognize, in the context of the additive white Gaussian noise channel, that to maximize
information transfer across a memoryless channel 
codewords must be
``noise-like'', i.e. resemble a typical sample of a memoryless random process with marginal
distribution that maximizes mutual information.
Specifically, in \cite[Section 25]{CS48} Shannon states:\footnote{In \cite{CS48} ``white noise" means white Gaussian noise.}
\begin{quote}
To approximate this limiting rate of transmission the transmitted signals must approximate, in statistical properties, a white noise.
\end{quote}
A general and formal statement of this property of optimal codes 
was put forward by Shamai and Verd\'{u}~\cite{SV97} who showed
that a capacity-achieving sequence of codes with \textit{vanishing} probability of
error must satisfy~\cite[Theorem 2]{SV97}
\begin{equation}\label{eq:re0}
	 {1\over n} D(P_{Y^n} || P_{Y^n}^*) \to 0\,,
\end{equation}
where $P_{Y^n}$ denotes the output distribution induced by the codebook (assuming
equiprobable codewords) and
\begin{equation}
	P_{Y^n}^* = P_Y^* \times\cdots\times P_Y^*
\end{equation}
is the $n$-th power of the single-letter capacity achieving output distribution $P_Y^*$ and
$D(\cdot||\cdot)$ is the relative entropy. Furthermore,~\cite{SV97} shows that (under 
regularity conditions) the empirical frequency of input letters (or sequential $k$-letter
blocks) inside the codebook approaches the capacity achieving input distribution (or its
$k$-th power) in the sense of vanishing relative entropy. 


In this paper, we extend the result in~\cite{SV97} to the case of \textit{non-vanishing} probability of error.
Studying this regime as opposed to \textit{vanishing} probability of error has recently proved to be
fruitful for the non-asymptotic characterization of the maximal achievable
rate~\cite{PPV08}. Although for the memoryless channels considered in this paper 
the $\epsilon$-capacity $C_\epsilon$ is independent of the probability of error
$\epsilon$, it does not immediately follow that a $C_\epsilon$-achieving code necessarily
satisfies the empirical distribution property~\eqref{eq:re0}. In fact, 
we will show that~\eqref{eq:re0} fails to be necessary
under the average probability of error criterion. 

To illustrate the delicacy of the question of approximating $P_{Y^n}$ with $P_{Y^n}^*$, consider a
good, capacity-achieving $k$-to-$n$ code for the binary symmetric channel (BSC) with crossover
probability $\delta <\frac12$ and capacity $C$. 
The probability of the codebook under $P_{Y^n}$ is larger than the probability that no errors occur: $(1-\delta)^n$. Under
$P_{Y^n}^*$ the probability of the codebook is $2^{k-n}$ 
\textemdash which is exponentially smaller asymptotically since for a
reliable code $k \leq n - n h(\delta) < \log 2(1-\delta)$). On the other hand,
consider a set $E$ consisting of a union of small Hamming balls surrounding
each codeword, whose radius $\approx \delta n$ is chosen such that $P_{Y^n}[E] = {1\over 2}$, say. Assuming that the code is decodable with small
probability of error, the union will be almost disjoint and hence $P_{Y^n}^*[E] \approx 2^{k-nC}$
\textemdash  the two becoming exponentially comparable (provided $k\approx nC$). Thus, for certain events,
$P_{Y^n}$ and $P_{Y^n}^*$ differ exponentially, while on other, less delicate, events
they behave similarly.  We will show that as long as the error probability is strictly less than one,
the normalized relative entropy in \eqref{eq:re0}
is upper bounded by the difference between capacity and code rate.

Studying the output distribution $P_{Y^n}$ also becomes important 
in the context of secure communication, where the output
due to the code is required to resemble white noise; and in the
problem of asynchronous communication where the output statistics of the code 
imposes limits on the quality of synchronization~\cite{TCW09}.
 For example, in a multi-terminal
communication problem, the channel output of one user may create interference for another. Assessing the
average impairment caused by such interference involves the analysis of the 
expectation of a certain function of
the channel output $\EE[F(Y^n)]$. We show that under certain regularity assumptions on $F$ not only
one can approximate the expectation of $F$ by substituting the unknown $P_{Y^n}$ with $P_{Y^n}^*$, as in
\begin{equation}\label{eq:approx}
	 \int F(y^n) dP_{Y^n} \approx \int F(y^n) dP_{Y^n}^*\,,
\end{equation}
but one can also prove that in fact the distribution of $F(Y^n)$ will be tightly
concentrated around its expectation. Thus, we are able to predict with overwhelming probability the
random value of $F(Y^n)$ without any knowledge of the code used to produce $Y^n$ (but assuming the
code is $\epsilon$-capacity-achieving).

Besides~\eqref{eq:re0} and~\eqref{eq:approx} we will show that 
\begin{enumerate}
\item the hypothesis testing problem
between $P_{Y^n}$ and $P_{Y^n}^*$ has zero Stein exponent;
\item a convenient inequality holds for the
conditional relative entropy for the channel output in terms of the cardinality of the employed
code;
\item codewords of good codes for the additive white 
Gaussian noise (AWGN) channel become more and more isotropically distributed 
(in the sense of evaluating quadratic forms) and resemble white Gaussian noise (in the sense of
$\ell_q$ norms) as the code
approaches the fundamental limits;
\item the output process $Y^n$ enjoys an asymptotic equipartition property.  
\end{enumerate} 
\par
Throughout the paper we will observe a number of
connections with the concentration of measure 
(isoperimetry) and optimal transportation, which were introduced into 
the information theory by the seminal works~\cite{AGK76,AD76,KM86}. Although some key results are stated
for general channels, most of the discussion is specialized to discrete memoryless
channels (DMC) (possibly with a (separable) input cost constraint) and to the AWGN channel.

The organization of the paper is as follows. Section~\ref{sec:notation} contains the main
definitions and notation. Section~\ref{sec:main} proves a sharp upper bound on the
relative entropy $D(P_{Y^n}||P_{Y^n}^*)$. In Section~\ref{sec:relent} we discuss various
implications of the bounds on relative entropy and in particular prove
approximation~\eqref{eq:approx}.
Section~\ref{sec:stein} considers the hypothesis testing problem of discriminating between
$P_{Y^n}$ and $P_{Y^n}^*$.  The asymptotic equipartition property of the channel output process
is established in Section~\ref{sec:aep}.
\ifincludenorms
Section~\ref{sec:awgn2} discusses results for the
quadratic forms and $\ell_p$ norms of the codewords of good Gaussian codes.
\fi

\section{Definitions and notation}\label{sec:notation}

\subsection{Codes and channels}
A random transformation $P_{Y|X}\colon\matx\to\maty$ is a Markov kernel acting between a pair of measurable
spaces.  An $(M, \epsilon)_{avg}$ code for the random transformation $P_{Y|X}$ is a
pair of random transformations $\sff \colon\{1,\ldots, M\}\to\matx$ and $\sfg \colon\maty\to\{1,\ldots,M\}$
such that
\begin{equation}\label{eq:epsavg}
	 \PP[\hat W \neq W] \le \epsilon\,,
\end{equation}
where in the underlying probability space $X=\sff(W)$ and $\hat W =\sfg(Y)$ with $W$
equiprobable on $\{1,\ldots,M\}$, and $W,X,Y,\hat W$ forming a Markov chain:
\begin{equation}\label{eq:probsp}
	 W \stackrel{\sff}{\to} X \stackrel{P_{Y|X}}{\to} Y \stackrel{\sfg}{\to} \hat W\,.
\end{equation}
In particular, we say that $P_X$ (resp., $P_Y$) is the input (resp., output) distribution induced by
the encoder $\sff$.
 An $(M, \epsilon)_{max}$ code is defined
similarly except that~\eqref{eq:epsavg} is replaced with the more stringent maximal
probability of error criterion:
\begin{equation}\label{eq:epsmax}
	 \max_{1 \le j \le M} \PP[\hat W \neq W | W=j] \le \epsilon\,.
\end{equation}
A code is  deterministic
if the encoder $\sff$ is a
functional (non-random) mapping.
We will frequently specify that a code is deterministic with the notation $(M, \epsilon)_{max, det}$ or $(M, \epsilon)_{avg, det}$.

A channel is a sequence of random transformations, $\{P_{Y^n|X^n},
n=1,\ldots\}$ indexed by the parameter $n$, referred to as the blocklength. 
An $(M, \epsilon)$ code for the $n$-th random
transformation is called an $(n, M, \epsilon)$ code, and the foregoing notation
specifying average/maximal error probability and deterministic encoder 
will also be applied to that case. The non-asymptotic fundamental limit of
communication is defined as\footnote{Additionally, one should also specify which
probability of error criterion,~\eqref{eq:epsavg} or~\eqref{eq:epsmax}, is used.}
\begin{equation}
	M^*(n, \epsilon) = \max \{M\colon \exists (n, M, \epsilon)\mbox{-code}\}\,.
\end{equation}
\subsection{Capacity-achieving output distribution}
To the three types of channels considered below we also associate a special sequence of output distributions
$P_{Y^n}^*$,
defined as the $n$-th power of a certain single-letter distribution $P_Y^*$:\footnote{For general
channels, the sequence $\{P_{Y^n}^*\}$ is required to satisfy a 
\textit{quasi-caod} property, see~\cite[Section IV]{PV11-relent}.}
\begin{equation}\label{eq:caod_spec}
	 P_{Y^n}^* \eqdef (P_Y^*)^n =  P_Y^* \times \cdots \times P_Y^*,
\end{equation}
where $P_Y^*$ is a distribution on the output alphabet defined as follows:
\begin{enumerate}
\item a DMC (without feedback) is built from a  single letter transformation
$P_{Y|X}\colon\matx \to \maty$ acting between finite spaces by extending the latter to all
$n\ge1$ in a memoryless way. Namely, the input space of the $n$-th random transformation
$P_{Y^n|X^n}$ is given by\footnote{To unify notation we denote the input
space as $\matx_n$ (instead of the more natural $\matx^n$) even in the absence of cost constraints.}
\begin{equation}
		\matx_n = \matx^n \eqdef \matx \times \ldots \times \matx
\end{equation}
and similarly for the output space $\maty^n = \maty \times \ldots\times \maty$, while the
transition kernel is set to be
\begin{equation}
	P_{Y^n |X^n}(y^n | x^n) = \prod_{j=1}^n P_{Y|X}(y_j | x_j)\,.
\end{equation}
	The capacity $C$ and $P_Y^*$, the unique capacity-achieving output distribution ({\it
caod}), are found by solving 
\begin{equation} \label{maxmi}
	C = \max_{P_X} I(X;Y)\,.
\end{equation}
\item a DMC with input constraint $(\mcost, P)$ is a generalization of the previous construction
with an additional restriction on the input space $\matx_n$:
\begin{equation}
	\matx_n = \left\{x^n\in\matx^n\colon \sum_{j=1}^n \mcost(x_j) \le nP \right\}
\end{equation}
	In this case the capacity $C$ and the caod $P_Y^*$ are found by restricting the maximization in \eqref{maxmi}
to those $P_X$ that satisfy
\begin{equation}
	\EE[\mcost(X)] \le P\,.
\end{equation}
\item the $AWGN(P)$ channel has an input space\footnote{For convenience we denote the 
elements of $\mreals^n$ as $\vect x$, $\vect y$ (for non-random
vectors) and $X^n, Y^n$ (for the random vectors).}
\begin{equation}
	\matx_n = \left\{\vect{x} \in \mreals^n\colon \norm{\vect x}_2 \le \sqrt{nP} \right\}
\end{equation}
the output space $\maty^n = \mreals^n$ and the transition kernel
\begin{equation}
	P_{Y^n|X^n = \vect x} = \matn(\vect x, \mathbf{I}_n)\,,
\end{equation}
	where $\matn(\vect x, \bSigma)$ denotes a (multidimensional) normal distribution with
mean $\vect x$ and covariance matrix $\bSigma$ and $\mathbf{I}_n$ -- is the $n\times n$
identity matrix. Then\footnote{As usual, all logarithms $\log$ and exponents $\exp$ are taken to
arbitrary fixed base, which also specifies the information units.}
	\begin{eqnarray} C &=& {1\over 2} \log (1+P)\\
		   P_Y^* &=& \matn(0, 1+P)\,.
	\end{eqnarray}
\end{enumerate}

As shown in~\cite{FT67,JK74} in all three cases $P_Y^*$ is unique and $P_{Y^n}^*$ satisfies the key
property: 
\begin{equation}\label{eq:caod_prop}
	 D(P_{Y^n|X^n = x} || P_{Y^n}^* ) \le n C \,,
\end{equation}
for all $x\in \matx_n$.  Since $ I ( U; V) = D (P_{V|U} \| Q | P_U ) - D (P_V \| Q )$,
Property~\eqref{eq:caod_prop} implies that for every input
distribution $P_{X^n}$ the induced output distribution $P_Y^n$ satisfies
\begin{eqnarray} 
  D(P_{Y^n} || P_{Y^n}^*) &\le& nC - I(X^n;Y^n)\,.\label{eq:sv97} \\
  P_{Y^n} &\ll& P_{Y^n}^* \label{eq:sv97a}\\
  P_{Y^n|X^n=x^n} &\ll& P_{Y^n}^*, \qquad\forall x^n\in\matx_n\,,\label{eq:sv97b}
\end{eqnarray}
As a consequence of~\eqref{eq:sv97b} the information density is well defined:
\begin{equation} \label{infodensity}
	\imath^*_{X^n; Y^n}(x^n;y^n) \eqdef \log {d P_{Y^n | X^n =x^n}\over dP_{Y^n}^*}(y^n)\,.
\end{equation}
Moreover, for every channel considered here there is a constant $a_1 > 0$ such
that\footnote{For discrete channels~\eqref{eq:caod_var} is shown, e.g., in~\cite[Appendix
E]{PPV08}.}
\begin{equation}\label{eq:caod_var}
	\sup_{x^n\in\matx_n} \Var \left[\imath^*_{X^n;Y^n}(X^n; Y^n)\,\middle|\, X^n = x^n\right] \le n a_1\,.
\end{equation}

In all three cases, the $\epsilon$-capacity $C_\epsilon$ equals $C$ for all $0<\epsilon<1$, i.e. 
\begin{equation}
	\log M^*(n, \epsilon) = nC + o(n)\,, \qquad n\to \infty\,.
\end{equation}
In fact, see~\cite{PPV08}
\begin{equation}
	\log M^*(n, \epsilon) = nC - \sqrt{nV} Q^{-1}(\epsilon) + O(\log n)\,, \qquad
n\to\infty\,,
\end{equation}
for any $0<\epsilon < {1\over 2}$,  a certain constant $V\ge 0$, called the channel dispersion, and $Q^{-1}$ is the inverse
of the standard complementary normal cdf.

\subsection{Good codes}
We introduce the following increasing 
degrees of optimality for sequences of $(n, M_n, \epsilon)$ codes. A code sequence is called:
\begin{enumerate}
\item \textit{$o(n)$-achieving} or $\epsilon$-capacity-achieving if
\begin{equation}\label{eq:capach}
	 {1\over n} \log M_n \to C\,. 
\end{equation}
\item \textit{$O(\sqrt{n})$-achieving} if
\begin{equation}
	\log M_n = nC + O(\sqrt{n})\,.
\end{equation}
\item \textit{$o(\sqrt{n})$-achieving} or dispersion-achieving if
\begin{equation}
	\log M_n = nC - \sqrt{nV} Q^{-1}(\epsilon) + o(\sqrt{n})\,.
\end{equation}
\item \textit{$O(\log n)$-achieving} if
\begin{equation}
	\log M_n = nC - \sqrt{nV} Q^{-1}(\epsilon) + O(\log n)\,.
\end{equation}
\end{enumerate}

\subsection{Binary hypothesis testing}
We also need to introduce the performance of an optimal
binary hypothesis test, which is one of the main tools in~\cite{PPV08}. Consider an
$\mata$-valued random variable $B$ 
which can take probability measures $P$ or $Q$.
A randomized test between those two distributions is defined by a random transformation
$P_{Z | B}\colon \mata \mapsto \{  0 ,  1 \}$ where $ 0$ indicates that the test chooses $Q$.
The best performance achievable among those randomized tests is given by\footnote{We
sometimes write summations over alphabets for simplicity of exposition. For arbitrary measurable
spaces $\beta_\alpha(P,Q)$ is defined by replacing the summation in~\eqref{eq:betadef} by an
expectation. }
\begin{equation} \label{eq:betadef}
	\beta_\alpha(P, Q) = 
				\min 
				\sum_{a \in \mata} Q  (a)  P_{Z | B} ({ 1} | a )\,,
	\end{equation}
where the minimum is over all probability distributions $P_{Z|B}$ satisfying
\begin{equation} P_{Z | B}\colon\\  \sum_{a \in \mata}  P  (a)  P_{Z | B} ({ 1} | a ) 
				 \ge \alpha\,.
\end{equation}
The minimum in~\eqref{eq:betadef} is guaranteed to be achieved by the Neyman-Pearson lemma. 
Thus, $\beta_\alpha(P, Q)$ gives the minimum 
probability of error under hypothesis $Q$ if the probability of error under hypothesis $P$
is no larger than $1 - \alpha$.

\section{Upper bound on the output relative entropy}\label{sec:main}

The main goal of this section is to establish (for each of the three types of memoryless channels
introduced in Section {sec:notation}) that
\begin{equation}\label{eq:re1_x}
	 D(P_{Y^n} || P_{Y^n}^*) \le nC - \log M_n + o(n)\,,
\end{equation}
where $P_{Y^n}$ is the sequence of output distributions induced by a sequence of $(n, M_n,
\epsilon)$ codes, and $o(n)$ depends on $\epsilon$.  Furthermore, for all channels except DMCs with zeros in the transition matrix $P_{Y|X}$, $o(n)$ in \eqref{eq:re1_x} can be replaced by $O ( \sqrt{n} )$.

We start by 
giving a one-shot converse due to Augustin \cite{UA66}
in Section~\ref{sec:sf}). Then, we prove~\eqref{eq:re1_x} for
DMCs in Section~\ref{sec:dmcboth} and for the AWGN
in Section~\ref{sec:awgn1}.

\subsection{Augustin's  converse}\label{sec:sf}

The following result  first appeared as part of the proofs
in~\cite[Satz 7.3 and 8.2]{UA66} by Augustin and formally stated in~\cite[Section 2]{RA82}.
Note that particularizing Theorem \ref{th:augustin} to a constant function $\rho$ recovers the 
nonasymptotic converse bound that can be derived from Wolfowitz's proof of the strong converse
\cite{JW57}.

\begin{theorem}[\cite{UA66,RA82}]\label{th:augustin} Consider a random transformation $P_{Y|X}$, a distribution $P_X$ induced by an 
$(M, \epsilon)_{max,det}$ code, a distribution $Q_Y$ on the output alphabet and a function
$\rho\colon\matx\to\mreals$. Then, provided the denominator is positive, 
\begin{equation}\label{eq:augustin}
	 M \le {\exp\{\EE[\rho(X)]\}\over
		\inf_{x} P_{Y|X=x}\left[ \log {dP_{Y|X=x}\over dQ_Y}(Y) \le \rho(x)\right]
- \epsilon}\,,
\end{equation}
 with the infimum taken over the support of $P_X$.
\end{theorem}
\begin{IEEEproof} Fix a $(M, \epsilon)_{max,det}$ code, $Q_Y$, and the function $\rho$. Denoting
 by $c_i$ the $i$-th codeword, we have 
\begin{equation}\label{eq:ag0}
	 Q_Y[\hat W(Y) = i] \ge \beta_{1-\epsilon}(P_{Y|X=c_i}, Q_Y)\,, \quad
i=1,\ldots,M\,,
\end{equation}
since $\hat W(Y)=i$ is a suboptimal test to decide between $P_{Y|X=c_i}$ and $Q_Y$, which achieves
error probability no larger than $\epsilon$ when $P_{Y|X=c_i}$ is true. Therefore,
\begin{eqnarray} {1\over M} &\ge& {1\over M} \sum_{i=1}^M \beta_{1-\epsilon}(P_{Y|X=c_i},
Q_Y)\label{eq:ag1}\\
	&\ge& {1\over M} \sum_{i=1}^M \left(P_{Y|X=c_i}\left[ \log {dP_{Y|X=c_i}\over
dQ_Y}(Y) \le \rho(c_i)\right] - \epsilon\right)\exp\{-\rho(c_i)\}\label{eq:ag2}\\
	&\ge& \left(\inf_x P_{Y|X=x}\left[ \log {dP_{Y|X=x}\over
dQ_Y}(Y) \le \rho(x)\right] - \epsilon\right){1\over M} \sum_{i=1}^M \exp\{-\rho(c_i)\}\label{eq:ag3}\\
	&\ge& \left(\inf_x P_{Y|X=x}\left[ \log {dP_{Y|X=x}\over
dQ_Y}(Y) \le \rho(x)\right] - \epsilon\right)\exp\{-\EE[\rho(X)]\}\label{eq:ag4}\,,
\end{eqnarray}
where~\eqref{eq:ag1} is by taking the arithmetic average of~\eqref{eq:ag0} over $i$,~\eqref{eq:ag4}
is by Jensen's inequality, and~\eqref{eq:ag2}
is by the standard estimate of $\beta_\alpha$, e.g.~\cite[(102)]{PPV08},
\begin{eqnarray}\label{eq:beta_est}
	 \beta_{1-\epsilon}(P, Q) 
	  \ge \left(\PP\left[ \log {dP\over dQ}(Z) \le \rho\right] - \epsilon \right) \exp\{-\rho\}\,,\label{eq:sf2}
\end{eqnarray}
with $Z$ distributed according to $P$.
\end{IEEEproof}

\begin{remark} Following an idea of Poor and Verd\'u~\cite{PV95} we may further 
strengthen Theorem~\ref{th:augustin} in the special case of $Q_Y=P_Y$: {\it The maximal probability of
error $\epsilon$ for any test of $M$ hypotheses $\{P_j, j=1,\ldots,M\}$ satisfies:
\begin{equation}\label{eq:augustin_pv}
	 \epsilon \ge \left(1 - {\exp \{\bar \rho\}\over M}\right) \inf_{1\le j \le M}
\PP\left[\imath_{W;Y}(W; Y) \le  \rho_j | W=j\right]\,,
\end{equation}
where the information density is as defined in \eqref{infodensity}, $\rho_j \in \mreals$ are arbitrary, $\bar \rho = {1\over M} \sum_{j=1}^M \rho_j$,
$W$ is equiprobable on $\{1,\ldots, M\}$ and $P_{Y|W=j}=P_j$.
}
Indeed, since $\imath_{W;Y}(a;b) \le \log M$ we get from~\cite[Lemma 35]{YP10}
\begin{equation}
	\exp\{\rho_j\} \beta_{1-\epsilon}(P_{Y|W=j}, P_Y) + \left(1 - {\exp\{\rho_j\}\over
M}\right) \PP\left[ \imath_{W;Y}(W; Y) > \rho_j | W=j\right] \ge 1-\epsilon\,.
\end{equation}
Multiplying by
$\exp\{-\rho_j\}$ and using resulting bound in place of~\eqref{eq:beta_est}
we repeat steps~\eqref{eq:ag1}-\eqref{eq:ag4} to obtain
\begin{equation}
	{1\over M} \inf_{1\le j \le M}
\PP\left[\imath_{W;Y}(W;Y) \le \rho_j | W=j\right] \ge \left(\inf_{1\le j \le M}
\PP\left[\imath_{W;Y}(W;Y) \le \rho_j | W=j\right] - \epsilon \right) \exp\{-\bar
\rho\}\,,
\end{equation}
which in turn is equivalent to~\eqref{eq:augustin_pv}.
\end{remark}

Choosing $\rho(x) = D(P_{Y|X=x}||Q_Y) + \Delta$ we can specialize
Theorem~\ref{th:augustin} in the
following convenient form:

\begin{theorem}\label{th:sf}
Consider a random transformation $P_{Y|X}$, a distribution $P_X$ induced by an $(M,
\epsilon)_{max,det}$ code and an auxiliary output distribution $Q_Y$. Assume that for all
$x\in \matx$ we have
\begin{equation}
	d(x) \eqdef D(P_{Y|X=x}||Q_Y) < \infty
\end{equation}
and 
\begin{equation}\label{eq:sf1}
	 \sup_x P_{Y|X=x}\left[ \log {dP_{Y|X=x}\over dQ_Y}(Y) \ge d(x) +
\Delta\right] \le \delta'\,,
\end{equation}
for some pair of constants $\Delta\ge 0$ and $0\le\delta'<1-\epsilon$. Then, we have
\begin{equation}\label{eq:sf}
	D(P_{Y|X} || Q_Y | P_X) \ge \log M - \Delta + \log (1-\epsilon - \delta')\,.
\end{equation}
\end{theorem}
\begin{remark} Note that~\eqref{eq:sf1} holding with a small $\delta'$ is a natural
non-asymptotic embodiment of information stability of the underlying channel, cf.~\cite[Section
IV]{PV11-relent}. 
\end{remark}

A simple way to estimate the upper deviations in~\eqref{eq:sf1} is by using Chebyshev's inequality.
As an example, we obtain
\begin{corollary}\label{th:sfvar} If in the conditions of Theorem~\ref{th:sf} we
replace~\eqref{eq:sf1} with
\begin{equation}\label{eq:sfvar_1}
	 \sup_x \Var\left[ \log {dP_{Y|X=x}\over dQ_Y}(Y) \middle| X=x\right] \le S_m 
\end{equation}
for some constant $S_m\ge0$, then we have
\begin{equation}\label{eq:sfvar}
	 D(P_{Y|X} || Q_Y | P_X) \ge \log M - \sqrt{2 S_m\over 1-\epsilon} +  \log
{1-\epsilon\over 2}\,.
\end{equation}
\end{corollary}
\subsection{DMC}\label{sec:dmcboth}
Notice that when $Q_Y$ is chosen to be a product distribution, such as $P_{Y^n}^*$, $\log
{dP_{Y|X=x}\over d Q_Y}$ becomes a sum of independent random variables. In particular,~\eqref{eq:caod_var} 
leads to a necessary and sufficient condition for~\eqref{eq:re0}:
\begin{theorem}\label{th:ideq} Consider a memoryless channel belonging to one of the three classes
in Section~\ref{sec:notation}. Then for any $0<\epsilon<1$ and any sequence of $(n, M_n, \epsilon)_{max,det}$
capacity-achieving codes we have
\begin{equation}\label{eq:ideq}
	 {1\over n} I(X^n; Y^n) \to C \, \iff \, {1\over n} D(P_{Y^n} || P_{Y^n}^*) \to 0 \,,
\end{equation}
where $X^n$ is the output of the encoder.
\end{theorem}
\begin{IEEEproof}
The direction $\Rightarrow$ is trivial from property~\eqref{eq:sv97} of $P_{Y^n}^*$. 
For the direction $\Leftarrow$ we only need to lower-bound $I(X^n;Y^n)$ since, asymptotically, it cannot exceed $nC$. To that end, we have from~\eqref{eq:caod_var} and Corollary~\ref{th:sfvar}:
\begin{equation}
	D(P_{Y^n | X^n} || P_{Y^n}^* | P_{X^n}) \ge \log M_n + O(\sqrt{n})\,.
\end{equation}
Then the conclusion follows from~\eqref{eq:capach} and the following identity applied with
$Q_{Y^n} = P_{Y^n}^*$:
\begin{equation}\label{eq:ideq2}
	 I(X^n; Y^n) = D(P_{Y^n|X^n} || Q_{Y^n} | P_{X^n}) - D(P_{Y^n} || Q_{Y^n})\,,
\end{equation}
which holds for all $Q_{Y^n}$ such that the unconditional relative entropy is finite.
\end{IEEEproof}

We remark that Theorem~\ref{th:ideq} can also be derived from a simple extension of the
Wolfowitz converse~\cite{JW57}
to an arbitrary output distribution $Q_{Y^n}$, e.g.~\cite[Theorem 10]{YP10}, and then choosing
$Q_{Y^n} = P_{Y^n}^*$. 
Note that Theorem~\ref{th:ideq} implies the result in~\cite{SV97} since the capacity-achieving codes
with vanishing error probability are a subclass of those considered in Theorem~\ref{th:ideq}.
\par
Fano's inequality only guarantees the left side of \eqref{eq:ideq} for code sequences with vanishing error probability.
If there was a strong converse showing that the left side of \eqref{eq:ideq} must hold for any sequence of $(n, M_n, \epsilon)$ codes,
then the desired result \eqref{eq:re0} would follow. In the absence of such a result we will consider three separate
cases in order to show  \eqref{eq:re0} , and, therefore, through Theorem \ref{th:ideq}, the left side of \eqref{eq:ideq}.
\subsubsection{DMC with $C_1 < \infty$}\label{sec:dmc1}
For a given DMC denote the parameter introduced by Burnashev~\cite{MVB76}
\begin{equation}
	C_1 = \max_{a,a'} D(P_{Y|X=a} || P_{Y|X=a'})\,.
\end{equation}
Note that $C_1 < \infty$ if and only if the transition matrix does not contain any zeros.
In this section we show~\eqref{eq:re1_x} for a (regular) class of DMCs with $C_1 < \infty$ by an
application of the main inequality~\eqref{eq:sf}. We also demonstrate that~\eqref{eq:re0} may not
hold for codes with non-deterministic encoders or unconstrained maximal probability of error.

\begin{theorem}\label{th:dmc1} Consider a DMC $P_{Y|X}$ with $C_1 < \infty$ and capacity $C>0$ (with or without an input
constraint).
Then for any $0\le\epsilon < 1$ there exists a constant $a = a(\epsilon)>0$ such that
 any $(n, M_n, \epsilon)_{max,det}$ code satisfies
\begin{equation}\label{eq:re1_dmc1}
	 D(P_{Y^n} || P_{Y^n}^*) \le nC - \log M_n + a\sqrt{n}\,,
\end{equation}
where $P_{Y^n}$ is the output distribution induced by the code.
In particular, for any capacity-achieving sequence of such codes we have
\begin{equation}\label{eq:dmc1}
	 {1\over n} D(P_{Y^n} || P_{Y^n}^*) \to 0\,, 
\end{equation}
\end{theorem}

\begin{IEEEproof} 
Fix $y^n, \bar y^n \in \maty^n$ which differ in the $j$-th letter only. Then, denoting 
$y_{\setminus j} = \{y_k, k\neq j\}$ we have
\begin{eqnarray} |\log P_{Y^n}(y^n) - \log P_{Y^n}(\bar y^n)|
		 &=&
		\left|\log {P_{Y_j | Y_{\setminus j}}(y_j | y_{\setminus j}) \over 
			P_{Y_j | Y_{\setminus j}}(\bar y_j | y_{\setminus j})} \right| \\
	&\le& \max_{a,b,b'} \log {P_{Y|X}(b|a)\over P_{Y|X}(b'|a)} \label{eq:dmc1_8b}\\
	&\eqdef& a_1 < \infty\,,\label{eq:dmc1_lip}
\end{eqnarray}
where~\eqref{eq:dmc1_8b} follows from
\begin{equation}
	P_{Y_j | Y_{\setminus j}}(b | y_{\setminus j}) = \sum_{a\in\matx} P_{Y|X}(b|a)
P_{X_j|Y_{\setminus j}}(a|y_{\setminus j})\,.
\end{equation}
Thus, the function $y^n \mapsto \log P_{Y^n}(y^n)$ is $a_1$-Lipschitz in Hamming metric on $\maty^n$. Its discrete gradient (absolute difference 
of values taken at consecutive integers) 
is bounded by $n |a_1|^2$ and
thus by the discrete Poincar\'e inequality (the variance of a function with countable support is upper bounded by (a multiple of) the second moment of its discrete gradient)~\cite[Theorem 4.1f]{BG99disc} we have
\begin{equation}\label{eq:dmc1_8}
	 \Var \left[\log P_{Y^n}(Y^n) \middle| X^n = x^n\right] \le n |a_1|^2\,.
\end{equation}
Therefore, for some $0<a_2<\infty$ and all $x^n \in \matx_n$ we have
\begin{eqnarray} 
\Var \left[\imath_{X^n;Y^n}(X^n; Y^n) \middle| X^n = x^n\right]	&\le& 2 \Var\left[\log P_{Y^n | X^n}(Y^n |X^n)  \middle| X^n = x^n\right] \nonumber\\
&& {} + 2 \Var
	\left[\log P_{Y^n}(Y^n) \middle| X^n = x^n\right] \label{eq:dmc1_7a}\\
&\le& 2n a_2 + 2n |a_1|^2\,,\label{eq:dmc1_7}
\end{eqnarray}
where
\eqref{eq:dmc1_7a}
follows from
\begin{equation} \label{jensencauchy}
\Var \left[ \sum_{i=1}^K Y_i \right] \leq K \sum_{i=1}^K \Var [  Y_i ]
\end{equation}
and~\eqref{eq:dmc1_7} follows from~\eqref{eq:dmc1_8} and the fact that the random variable in the
first variance in~\eqref{eq:dmc1_7a} is a sum of $n$
independent terms. Applying Corollary~\ref{th:sfvar} with
$S_m = 2na_2 + 2n |a_1|^2$ and $Q_Y = P_{Y^n}$ we obtain:
\begin{equation}\label{eq:dmc1_12}
	 D(P_{Y^n|X^n} || P_{Y^n} | P_{X^n}) \ge \log M_n + O(\sqrt{n})\,.
\end{equation}
We can now complete the proof:
\begin{align} 
	D(P_{Y^n} || P_{Y^n}^*)
	 &= D(P_{Y^n|X^n} || P_{Y^n}^* | P_{X^n}) -  
	D(P_{Y^n|X^n} || P_{Y^n} | P_{X^n})\\
		&\le nC - D(P_{Y^n|X^n} || P_{Y^n} | P_{X^n})\label{eq:dmc1_9}\\
		&\le nC - \log M_n + O(\sqrt{n}) \label{eq:dmc1_10}
\end{align}
where~\eqref{eq:dmc1_9} is because $P_{Y^n}^*$ satisfies~\eqref{eq:caod_prop}
and~\eqref{eq:dmc1_10} follows from~\eqref{eq:dmc1_12}. This completes the proof
of~\eqref{eq:re1_dmc1}.
\end{IEEEproof}
\begin{remark} 
As we will see in Section~\ref{sec:empir},~\eqref{eq:dmc1}  implies
\begin{equation}\label{eq:dmc1_entest}
	 H(Y^n) = n H(Y^*) + o(n) 
\end{equation}
(by~\eqref{eq:dconv_avg} applied to $f(y) = \log P_Y^*(y) $).
Note also that traditional combinatorial methods, e.g.~\cite{CK2}, are not helpful
in dealing with quantities like $H(Y^n)$, $D(P_{Y^n} || P_{Y^n}^*)$ or $P_{Y^n}$-expectations of
functions that are not of the form of cumulative average.
\end{remark}

\begin{remark} Note that any $(n, M, \epsilon)$ code is also an $(n, M, \epsilon')$ code for all
	$\epsilon'\ge\epsilon$. Thus $a(\epsilon)$, the constant in~\eqref{eq:re1_dmc1}, is a non-decreasing function of $\epsilon$. In
	particular,~\eqref{eq:re1_dmc1} holds uniformly in $\epsilon$ on compact subsets of $[0,1)$. 
In their follow-up to the present paper, Raginsky and Sason \cite{raginskyrefined} use McDiarmid's inequality
to derive a tighter estimate for $a$.
\end{remark}

\begin{remark}\label{rm:avg} \eqref{eq:dmc1} need not hold if the maximal
probability of error is replaced with the average or if the encoder is allowed to be random. 
Indeed, for any $0<\epsilon < 1$ we construct a sequence of $(n, M_n, \epsilon)_{avg}$
capacity-achieving codes which do not satisfy~\eqref{eq:dmc1} can be constructed as follows. 
Consider a sequence of $(n, M_n',
\epsilon_n')_{max,det}$ codes with $\epsilon_n'\to0$ and
\begin{equation}\label{eq:dmc1_2b}
	{1\over n} \log M_n' \to C\,.
\end{equation}
For all $n$ such that $\epsilon'_n < {1\over 2}$ this code cannot have repeated codewords
and we can additionally assume (perhaps by reducing $M_n'$ by one) that there is no
codeword equal to $(x_0, \ldots, x_0)\in \matx_n$, where $x_0$ is some fixed letter in
$\matx$ such that
\begin{equation}\label{eq:dmc1_2}
	D(P_{Y|X=x_0} || P_Y^*) > 0
\end{equation}
(the existence of such $x_0$ relies on the assumption $C>0$). Denote the output distribution
induced by this code by $P_{Y^n}'$.
Next, extend this code by adding ${\epsilon - \epsilon_n'\over 1-\epsilon} M_n'$ identical codewords:
$(x_0, \ldots, x_0)\in \matx_n$. Then the minimal average probability of 
error achievable with the extended codebook of size 
\begin{equation}
	M_n \eqdef {1-\epsilon_n\over 1-\epsilon} M_n'
\end{equation}
 is easily seen to be not larger than $\epsilon$. 
Denote the output distribution induced by the extended code
by $P_{Y^n}$ and define a binary random variable
\begin{equation}
	S = 1\{X^n = (x_0, \ldots, x_0) \}
\end{equation}
with distribution
\begin{equation}
	P_S(1) = 1-P_S(0) = {\epsilon - \epsilon_n'\over 1-\epsilon_n'}\,.
\end{equation}
which satisfies $P_S(1) \to \epsilon$.
We have then
\begin{align} D(P_{Y^n} || P_{Y^n}^*)
	 &= D(P_{Y^n | S} || P_{Y^n}^* | P_S) - I(S;Y^n)\label{eq:dmc1_3} \\
	&\ge D(P_{Y^n | S} || P_{Y^n}^* | P_S) - \log 2 \label{eq:dmc1_4}\\
	&= n D(P_{Y|X=x_0} || P_Y^*) P_S(1) + D(P_{Y^n}' || P_{Y^n}^*) P_S(0) - \log 2\label{eq:dmc1_5} \\
	&= n D(P_{Y|X=x_0} || P_Y^*) P_S(1) + o(n)\label{eq:dmc1_6}\,, 
\end{align}
where~\eqref{eq:dmc1_3} is by~\eqref{eq:ideq2},~\eqref{eq:dmc1_4} follows since $S$ is binary,~\eqref{eq:dmc1_5} is by noticing that $P_{Y^n|S=0} = P_{Y^n}'$, and~\eqref{eq:dmc1_6} is
by~\eqref{eq:dmc1}. It is clear that~\eqref{eq:dmc1_2} and~\eqref{eq:dmc1_6} show the impossibility
of~\eqref{eq:dmc1} for this code.

Similarly, one shows that~\eqref{eq:dmc1} cannot hold if the assumption of the
deterministic encoder is dropped. Indeed, then we can again take the very same $(n, M_n', \epsilon_n')$ code and make its
encoder randomized so that with probability ${\epsilon - \epsilon_n'\over 1-\epsilon_n'}$
it outputs $(x_0, \ldots, x_0)\in\matx_n$ and otherwise it outputs the original codeword. The same
analysis shows that~\eqref{eq:dmc1_6} holds again and thus~\eqref{eq:dmc1} fails.

The counterexamples constructed above can also be used to demonstrate that in
Theorem~\ref{th:sf} (and hence Theorem~\ref{th:augustin}) the assumptions of maximal probability of error and deterministic encoders are
not superfluous, contrary to what is claimed by Ahlswede~\cite[Remark 1]{RA82}.
\end{remark}

\subsubsection{DMC with $C_1 = \infty$}\label{sec:dmc2}
Next, we show an estimate for $D(P_{Y^n}||P_{Y^n}^*)$ differing by a $\log^{3\over 2} n$ factor
from~\eqref{eq:re1_x} for the DMCs with $C_1 = \infty$.
\begin{theorem}\label{th:dmc}
 For any DMC $P_{Y|X}$ with capacity $C>0$ (with or without input constraints), $C_1 = \infty$, and
 $0 \le \epsilon < 1$  there exists a constant $b>0$ with the property
that for any sequence of $(n, M_n, \epsilon)_{max,det}$ codes we have for all $n\ge 1$
\begin{equation}\label{eq:dmc0}
	 D(P_{Y^n} || P_{Y^n}^*) \le nC - \log M_n + b \sqrt{n} \log^{3\over 2} n\,.
\end{equation}
In particular, for any such sequence achieving capacity we have
\begin{equation}\label{eq:dmc}
	 {1\over n} D(P_{Y^n} || P_{Y^n}^*) \to 0\,.
\end{equation}
\end{theorem}

\begin{IEEEproof} 
Let $c_i$ and $D_i, i =1,\ldots M_n$ denote the codewords and the 
decoding regions of the code. Denote the sequence
\begin{equation}
	\ell_n = b_1 \sqrt{n \log n}
\end{equation}
with $b_1 > 0$ to be further constrained shortly.
According to the isoperimetric inequality for Hamming space~\cite[Corollary
I.5.3]{CK2}, there is a constant $a>0$ such that for every $i=1,\ldots,M_n$
\begin{eqnarray} 
	1 - P_{Y^n | X^n = c_i}[\Gamma^{\ell_n} D_i] &\le& Q\left( Q^{-1}(\epsilon) +
{\ell_n\over \sqrt{n}} a\right)\\
		&\le& \exp\left\{- b_2 {\ell_n^2 \over n}  \right\} \\
		&=& n^{-b_2}\\
		&\le& {1\over n} \label{eq:dmc_0b}\,,
\end{eqnarray}
where the $\ell$-blowup of $D$ is defined as
\begin{equation}
	\Gamma^\ell D = \left\{\bar{y}^n \in \maty^n: \exists y^n \in D\mbox{~s.t.~} |\{j: y_j \neq \bar{y}_j\}| \le \ell
\right\}
\end{equation}
denotes the $\ell$-th Hamming neighborhood of a set $D$ and
 we assumed that $b_1$ was chosen large enough so there is $b_2\ge1$ satisfying~\eqref{eq:dmc_0b}.

Let 
\begin{equation}\label{eq:dmc_0}
	 M_n' = {M_n \over n{n \choose \ell_n} |\maty|^{\ell_n}}
\end{equation}
and consider a subcode $F = (F_1,\ldots, F_{M_n'})$, where $F_i \in \matc = \{ c_1 , \ldots, c_M \}$ and note that we allow
repetition of codewords. 
Then for every possible choice of the subcode $F$ we denote by $P_{X^n(F)}$ and $P_{Y^n(F)}$ the input/output distribution
induced by $F$, so that for example:
\begin{equation}
	P_{Y^n(F)} = {1\over M_n'} \sum_{j=1}^{M_n'} P_{Y^n|X^n = F_j}\,.
\end{equation}
We aim to apply the random coding argument over all equally likely $M_n^{M_n'}$ choices of a subcode $F$. 
Random coding among subcodes was originally invoked in~\cite{AD76} to demonstrate the existence of a good subcode.
The expected (over the choice of $F$) induced output distribution is
\begin{eqnarray} 
\EE[P_Y^n(F)] &\eqdef& {1\over M_n^{M_n'}} \sum_{F_1\in \matc} \cdots \sum_{F_{M_n'} \in \matc}
P_{Y^n(F)}\\
&=&
{1\over M_n^{M_n'}} {1\over M_n'} \sum_{j=1}^{M_n'} \sum_{F_1\in \matc} \cdots \sum_{F_{M_n'} \in \matc} 
P_{Y^n|X^n = F_j} \\
&=&
{M_n^{M_n'-1}\over M_n^{M_n'}}  \sum_{c \in \matc} P_{Y^n|X^n = c} \\
	&=& P_{Y^n}\,.\label{eq:dmc_0a}
\end{eqnarray}

Next, for every $F$ we denote by $\epsilon'(F)$ the minimal possible average probability
of error achieved by an appropriately chosen decoder. With this notation we have, for
every possible value of $F$:
\begin{align} D(P_{Y^n(F)} || P_{Y^n}^*)
	 &= 
	D(P_{Y^n|X^n} || P_{Y^n}^*| P_{X^n(F)}) - I(X^n(F); Y^n(F)) \label{eq:dmc_1}\\
	&\le nC - I(X^n(F); Y^n(F))\label{eq:dmc_2}\\
	&\le nC - (1-\epsilon'(F)) \log M_n' + \log 2\label{eq:dmc_3}\\
	&\le nC - \log M_n' + n \epsilon'(F) \log |\matx| + \log 2\label{eq:dmc_5}\\
	&\le nC - \log M_n + n\epsilon'(F) \log |\matx| + b_3 \sqrt{n} \log^{3\over 2} n\label{eq:dmc_6}
\end{align}
where~\eqref{eq:dmc_1} is by~\eqref{eq:ideq2},~\eqref{eq:dmc_2} is
by~\eqref{eq:caod_prop},~\eqref{eq:dmc_3} is by Fano's inequality,~\eqref{eq:dmc_5} is
because $\log M_n' \le n \log |\matx|$ and~\eqref{eq:dmc_6} holds for some $b_3>0$ 
by the choice of $M_n'$ in~\eqref{eq:dmc_0} and by
\begin{equation}
	\log {n \choose \ell_n} \le \ell_n \log n \,.
\end{equation}

Taking the expectation of both sides of~\eqref{eq:dmc_6}, applying convexity of relative
entropy and~\eqref{eq:dmc_0a} we get
\begin{align} 
	D(P_{Y^n} || P_{Y^n}^*)
	&\le nC - \log M_n + n\EE[\epsilon'(F)] \log |\matx| + b_3 \sqrt{n}
\log^{3\over 2} n\,. 
\end{align}
Accordingly, it remains to show that 
\begin{equation}\label{eq:dmc_10}
	 n \EE[\epsilon'(F)]  \le 2\,. 
\end{equation}
To that end, for every subcode $F$ define the suboptimal randomized decoder: 
\begin{equation}
	\hat W(y) = F_j \qquad \forall F_j\in L(y,F) \quad \mbox{(with probability ${1\over |L(y,
F)|})$}\,,
\end{equation}
where $L(y, F)$ is a list of those indices $i\in F$ for which $y\in
\Gamma^{\ell_n}D_i$. Since the transmitted codeword $F_W$ is equiprobable on $F$,  averaging over the selection of $F$ we have
\begin{equation}\label{eq:dmc_7a}
	 \EE[|L(Y^n, F) | \,|\, F_W \in L(Y^n,F)] \le 1 + {{n \choose \ell_n} |\maty|^{\ell_n}\over M_n}
(M_n' -1)\,,
\end{equation}
because each $y\in
\maty^n$ can belong to at most ${n\choose \ell_n} |\maty|^{\ell_n}$ enlarged decoding
regions $\Gamma^{\ell_n} D_i$ and  each $F_j$ is chosen independently and
equiprobably among all possible $M_n$ alternatives. The average (over random decoder, $F$, and channel) probability of error for
 can be upper-bounded as 
\begin{eqnarray} 
	\EE[\epsilon'(F)] \label{review15}
	 &=& \PP[ F_W \not \in L(Y^n, F) ] 
	 + \EE\left[ {|L(Y^n, F)| - 1\over |L(Y^n, F)|} 1\{F_W \in L(Y^n, F)\}\right]\\
	&\le& \PP[ F_W \not \in L(Y^n, F) ] + {{n \choose \ell_n} |\maty|^{\ell_n} M_n'\over M_n}\label{eq:dmc_8}\\
	&\le& {1\over n} + {{n \choose \ell_n} |\maty|^{\ell_n} M_n'\over M_n}\label{eq:dmc_7}\\
	&\le& {2\over n}\,,\label{eq:dmc_9}
\end{eqnarray}
where  \eqref{review15} reflects the fact that a correct decision requires that the true codeword not only belong to
$L(Y^n, F)$ but that it be the one chosen from the list;~\eqref{eq:dmc_8} is by Jensen's inequality applied to $x-1\over x$
and~\eqref{eq:dmc_7a};~\eqref{eq:dmc_7} is by~\eqref{eq:dmc_0b}; and~\eqref{eq:dmc_9} is by~\eqref{eq:dmc_0}.
Since~\eqref{eq:dmc_9} also serves as an upper bound to $\EE[\epsilon'(F)]$ the proof
of~\eqref{eq:dmc_10} is complete.
\end{IEEEproof}
\begin{remark}
 Claim~\eqref{eq:dmc} fails to hold if either the maximal
probability of error is replaced with the average, or if we allow the encoder to be stochastic.
Counterexamples are constructed exactly as in Remark~\ref{rm:avg}.
\end{remark}
\begin{remark}
Raginsky and Sason \cite{raginskyrefined} give a sharpened version of \eqref{eq:dmc0}
with explicitly computed constants but with the same $O( \sqrt{ n \log^3 n } )$ remainder term behavior.
\end{remark}
\subsection{Gaussian channel}\label{sec:awgn1}

\begin{theorem}\label{th:awgn1} For any $0<\epsilon<1$ and $P>0$ there exists $a = a(\epsilon, P)>0$ such 
that the output distribution $P_{Y^n}$ of any $(n, M_n, \epsilon)_{max,det}$ code for the
$AWGN(P)$ channel  satisfies\footnote{More precisely, our proof yields a bound  
	$nC - \log M + \sqrt{6n(3+4P)} \log e + \log{2\over
1-\epsilon}$.}
\begin{equation}\label{eq:re1_awgn1}
	 D(P_{Y^n} || P_{Y^n}^*) \le nC - \log M + a\sqrt{n}\,,
\end{equation}
where $P_{Y^n}^* = \matn(0, 1+P)^n$.
In particular for any capacity-achieving sequence of such codes we have
\begin{equation}\label{eq:awgn1}
	 {1\over n} D(P_{Y^n} || P_{Y^n}^*) \to 0\,.
\end{equation}
\end{theorem}

\begin{IEEEproof} 
Denote by $p_{Y^n|X^n=\vect x}$ and $p_{Y^n}$ the densities of
$P_{Y^n|X^n=\vect x}$ and $P_{Y^n}$, respectively.
The argument proceeds step by step as in the proof of Theorem~\ref{th:dmc1}
with~\eqref{eq:awgn1_8} taking the place of~\eqref{eq:dmc1_8} and recalling that
property~\eqref{eq:caod_prop} holds for the AWGN channel too.
Therefore, the objective is to show
\begin{equation}\label{eq:awgn1_8}
	 \Var[ \log p_{Y^n}(Y^n) \,|\, X^n ] \le a_1 n
\end{equation}
for some $a_1 > 0$. 
Poincar\'e's inequality for the Gaussian measure, e.g.~\cite[(2.16)]{MLxx}
states that if $Y$ is an $N$-dimensional Gaussian measure, then
\begin{equation} \label{reviw160}
	\Var[ f (Y) ] \le \EE[\norm{\nabla f (Y)}^2 ]
\end{equation}
Since conditioned on $X^n$, the random vector $Y^n$ is Gaussian, the Poincar\'e inequality
ensures that the left side of \eqref{eq:awgn1_8} is bounded by
\begin{equation} \label{reviw16}
	\Var[ \log p_{Y^n}(Y^n) \,|\, X^n ] \le \EE[\norm{\nabla \log p_{Y^n}}^2 \, |\, X^n]
\end{equation}
Therefore, the reminder of the proof is devoted to showing that the right side of \eqref{reviw16}
is bounded by $a_1 n$ for some $a_1 >0$.
An elementary computation shows
\begin{eqnarray}
	\nabla \log p_{Y^n}(\vect y) &=&  {\log e\over p_{Y^n}(\vect y)} \nabla p_{Y^n}(\vect y) \\
		&=& {\log e\over p_{Y^n}(\vect y)} 
			\sum_{j=1}^M {1\over M (2\pi)^{n\over 2}} \nabla e^{-{1\over 2}||\vect y - \vect
c_j||^2} \\
		&=& {\log e\over p_{Y^n}(\vect y)} \sum_{j=1}^M {1\over M(2\pi)^{n\over 2}}
			(\vect c_j - \vect y) e^{-{1\over 2}||\vect y - \vect c_j||^2} \\
		&=&
		(\EE[X^n | Y^n = \vect y] - \vect y) \log e\,.
\end{eqnarray}
For convenience denote
\begin{equation}
	\hat X^n = \EE[X^n | Y^n]
\end{equation}
and notice that since $\norm{X^n} \le \sqrt{nP}$ we have also
\begin{equation}\label{eq:awgn1_0}
	 \norm{\hat X^n} \le \sqrt{nP}\,.
\end{equation}
Then 
\begin{eqnarray}
 {1\over \log^2 e}\EE[\norm{\nabla \log p_{Y^n}(Y^n)}^2 \,|\, X^n]
	&=& 
	\EE\left[ \norm{Y^n - \hat X^n}^2 \,\middle|\, X^n\right] \label{eq:awgn1_1}\\
	&\le& 2\EE\left[ \norm{Y^n}^2 \,\middle|\, X^n\right] + 2 \EE\left[\norm{\hat
X^n}^2 \,\middle|\, X^n\right]\label{eq:awgn1_2}\\
	&\le& 2\EE\left[ \norm{Y^n}^2 \,\middle|\, X^n\right] + 2 nP\label{eq:awgn1_3}\\
	 &=& 2\EE\left[ \norm{X^n + Z^n}^2 \,\middle|\, X^n\right] + 2nP\label{eq:awgn1_4}\\
	&\le& 4 \norm{X^n}^2 + 4n + 2nP\label{eq:awgn1_5}\\
	&\le& (6P + 4)n\,,\label{eq:awgn1_6}
\end{eqnarray}
where~\eqref{eq:awgn1_2} is by 
\begin{equation}\label{eq:awgn1_7}
	 \norm{\vect a+ \vect b}^2 \le 2\norm{\vect a}^2 + 2 \norm{\vect b}^2\,,
\end{equation}
\eqref{eq:awgn1_3} is by~\eqref{eq:awgn1_0}, in~\eqref{eq:awgn1_4} we introduced $Z^n\sim
\matn(0, \mathbf{I}_n)$ which is independent of $X^n$,~\eqref{eq:awgn1_5} is
by~\eqref{eq:awgn1_7} and~\eqref{eq:awgn1_6} is by the power-constraint imposed on the codebook.
In view of \eqref{reviw16}, we have succeeded in identifying a constant $a_1$ such that  \eqref{eq:awgn1_8} holds.
\end{IEEEproof}

\begin{remark} \eqref{eq:awgn1} need not hold if the maximal
probability of error is replaced with the average or if the encoder is allowed to be stochastic. 
Counterexamples are constructed similarly to those for Remark~\ref{rm:avg} with $x_0 = 0$.
Note also that Theorem~\ref{th:awgn1} need not hold if the power-constraint is
in the average-over-the-codebook sense; see~\cite[Section 4.3.3]{YP10}.
\end{remark}

\section{Implications}\label{sec:relent}

We have shown that there is a constant $a=a(\epsilon)$ independent of $n$ and $M$ such that
\begin{equation}\label{eq:re1}
	 D(P_{Y^n} || P_{Y^n}^*) \le nC - \log M + a\sqrt{n}\,,
\end{equation}
where $P_{Y^n}$ is the output distribution induced by an arbitrary $(n, M, \epsilon)_{max,det}$ code.
 Therefore,  any $(n, M, \epsilon)_{max,det}$ necessarily satisfies
\begin{equation}\label{eq:logm_up}
	\log M \le nC + a(\epsilon) \sqrt{n}
\end{equation}
as is classically known~\cite{JW62}. 
In particular,~\eqref{eq:re1} implies that any $\epsilon$-capacity-achieving code must satisfy~\eqref{eq:re0}.
In this section we discuss this and other implications of this result, such as:
\begin{enumerate}
\item \eqref{eq:re1} implies that the empirical marginal output distribution
\begin{equation}\label{eq:emp_output}
	\bar P_n \eqdef {1\over n} \sum_{j=1}^n P_{Y_i}
\end{equation}
	converges to $P_Y^*$ in a strong sense (Section~\ref{sec:empir});
\item \eqref{eq:re1} guarantees estimates of the precision in the approximation~\eqref{eq:approx}
(Sections~\ref{sec:approx_mean} and~\ref{sec:approx_mean2}),
\item \eqref{eq:re1} provides estimates for the deviations of $f(Y^n)$ from its
average (Sections~\ref{sec:approx_conc}).
\item relation to optimal transportation (Section~\ref{sec:opttrans}),
\item implications of~\eqref{eq:re0} for the empirical input distribution of the code
(Sections~\ref{sec:dmcinp} and~\ref{sec:awgninp}).
\end{enumerate}

\subsection{Empirical distributions and empirical averages}\label{sec:empir}
Considering the empirical marginal distributions, the convexity of relative entropy
and~\eqref{eq:re0} result in
\begin{equation}\label{eq:dconv1}
	 D(\bar P_n || P_Y^*) \le {1\over n} D(P_{Y^n} || P_{Y^n}^*) \to 0\,,
\end{equation}
where $\bar P_n$ is the empirical marginal output distribution~\eqref{eq:emp_output}.

More generally, we have~\cite[(41)]{SV97}
\begin{equation}\label{eq:dconvk}
	 D(\bar P_n^{(k)} || P_{Y^k}^*) \le {k\over n-k+1}D(P_{Y^n} || P_{Y^n}^*) \to 0\,, 
\end{equation}
where $\bar P_n^{(k)}$ is a $k$-th order empirical output distribution
\begin{equation}\label{eq:dk_def}
	 \bar P_n^{(k)} = {1\over n-k+1} \sum_{j=1}^{n-k+1} P_{Y_j^{j+k-1}} \,.
\end{equation}

Knowing that a sequence of distributions $P_n$ converges in relative entropy to a
distribution $P$, i.e.
\begin{equation}\label{eq:relent_conv}
	 D(P_n || P) \to 0 
\end{equation}
implies convergence properties for the expectations of functions
\begin{equation}\label{eq:fconv}
	 \int f dP_n \to \int f dP
\end{equation}

\begin{enumerate}
\item For bounded functions, \eqref{eq:fconv} follows from the Csisz\'ar-Kemperman-Kullback-Pinsker inequality (e.g.~\cite{IC67}):
\begin{equation}\label{eq:pinsker}
	|| P_n - P ||_{TV}^2 \le {1\over 2 \log e} D(P_n || P)\,,
\end{equation}
where
\begin{equation}\label{eq:tv}
	 \norm{P - Q}_{TV} \eqdef \sup_A |P(A) - Q(A)| 
\end{equation} 

\item For unbounded $f$, \eqref{eq:fconv} holds as long as $f$ satisfies Cramer's condition under $P$, i.e.
\begin{equation}
	\int e^{t f} dP <\infty
\end{equation}
for all $t$ in some neighborhood of $0$; see~\cite[Lemma 3.1]{IC75}.
\end{enumerate}

Together~\eqref{eq:fconv} and~\eqref{eq:dconv1} show that for a wide class of functions
$f\colon\maty\to\mreals$  empirical averages  over distributions induced by good codes converge
to the average over the capacity achieving output distribution (caod):
\begin{equation}\label{eq:dconv_avg}
	\EE\left[ {1\over n}\sum_{j=1}^n f(Y_j) \right] \to \int f dP_Y^*\,.
\end{equation}
From~\eqref{eq:dconvk} a similar conclusion holds for $k$-th order empirical averages.

\subsection{Averages of functions of $Y^n$}\label{sec:approx_mean}
To go beyond empirical averages, we need to provide some definitions and properties (see \cite{MLxx})

\begin{definition} The function $F\colon\maty^n \to \mreals$ is called $(b, c)$-concentrated with respect
to measure $\mu$ on $\maty^n$ if for all $t\in\mreals$
\begin{equation}\label{eq:fcc}
	 \int \exp\{t (F(Y^n) - \bar F)\} d\mu \le b \exp\{ct^2\}\,, \qquad \bar F = \int F d\mu\,.
\end{equation}
A function $F$ is called $(b, c)$-concentrated for the channel if it is $(b,c)$-concentrated with respect to
every $P_{Y^n|X^n = x}$ and $P_{Y^n}^*$ and all $n$.
\end{definition}

A couple of simple properties of $(b,c)$-concentrated functions:
\begin{enumerate}
\item Gaussian concentration around the mean:
\begin{equation}
	\PP[ | F(Y^n) - \EE[F(Y^n)] | > t] \le b \exp\left\{-{t^2 \over 4c}\right\}\,.
\end{equation}
\item Small variance:
\begin{eqnarray} \Var[F(Y^n)] &=& \int_0^\infty \PP[ | F(Y^n) - \EE[F(Y^n)] |^2 > t] dt\\
			&\le& \int_0^\infty \min\left\{b \exp\left\{-{t \over 4c}\right\}, 1\right\} dt\\
			&=& 4c \log(2be)\,.\label{eq:re_var}
\end{eqnarray}
\end{enumerate}

Some examples of concentrated functions include:
\begin{itemize}
\item A bounded function $F$ with $\norm{F}_\infty \le A$ is $(\exp\{A^2 (4c)^{-1}\}, c)$-concentrated for
any $c$ and any measure $\mu$. Moreover, for a fixed $\mu$ and a sufficiently large $c$
any bounded function is $(1, c)$-concentrated.
\item If $F$ is $(b,c)$-concentrated then $\lambda F$ is $(b,\lambda^2
c)$-concentrated.
\item Let $f\colon\maty\to\mreals$ be $(1,c)$-concentrated with respect to $\mu$. Then so is 
\begin{equation}\label{eq:ex2}
	 F(y^n) = {1\over \sqrt{n}}\sum_{j=1}^n f(y_j)  
\end{equation}
with respect to $\mu^n$. In particular,
any $F$ defined in this way from a bounded $f$ is $(1,c)$-concentrated for a memoryless
channel (for a sufficiently large $c$ independent of $n$).
\item If $\mu = \matn(0, 1)^n$ and $F$ is a Lipschitz function on $\mreals^n$ with Lipschitz constant
$\norm{F}_{Lip}$ then $F$ is $(1, {\norm{F}^2_{Lip}\over
2 \log e})$-concentrated with respect to $\mu$, e.g.~\cite[Proposition 2.1]{ML94}:
\begin{equation}\label{eq:conc_gauss}
	 \int_{\mreals^n} \exp\{ t(F(y^n) - \bar F) \} d\mu(y^n) \le 
		\exp\left\{ {\norm{F}_{Lip}^2 \over 2 \log e} t^2 \right\}\,. 
\end{equation}
 Therefore any
Lipschitz function is $(1, {(1+P)\norm{F}^2_{Lip}\over 2\log e})$-concentrated for the AWGN channel.
\item For discrete $\maty^n$ endowed with the Hamming distance
\begin{equation}
	d(y^n, z^n) = |\{i\colon y_i \neq z_i\}|
\end{equation}
define Lipschitz functions in the usual way. In this case, a simpler criterion is: 
$F\colon\maty^n\to\mreals$ is Lipschitz with constant $\ell$ if and only if
\begin{equation}
   \max_{y^n, b, j}
	|F(y_1, \ldots, y_j, \ldots, y_n) - F(y_1, \ldots, b, \ldots, y_n)| \le \ell\,.
\end{equation}
Let $\mu$ be any product probability measure $P_1 \times \ldots \times P_n$ on $\maty^n$, then the 
standard Azuma-Hoeffding estimate shows that
\begin{equation}\label{eq:conc_hamming}
	 \sum_{y^n \in \maty^n} \exp\{ t(F(y^n) - \bar F) \} \mu(y^n) \le 
		\exp\left\{ {n \norm{F}_{Lip}^2 \over 2 \log e} t^2 \right\} 
\end{equation}
and thus any Lipschitz function $F$ is $(1, {n \norm{F}_{Lip}^2 \over 2 \log e})$-concentrated with
respect to any product measure on $\maty^n$.

Note that unlike the Gaussian case, the constant of concentration $c$ worsens linearly with dimension
$n$. Generally, this growth cannot be avoided as shown by the coefficient $1\over \sqrt{n}$ in the 
exact solution of the Hamming
isoperimetric problem~\cite{LH66}. At the same time, this growth does not
mean that~\eqref{eq:conc_hamming} is ``weaker'' than~\eqref{eq:conc_gauss}; for example, $F=\sum_{j=1}^n
\phi(y_j)$ has Lipschitz constant $O(\sqrt{n})$ in Euclidean space and $O(1)$ in Hamming. 
However, for convex functions the concentration~\eqref{eq:conc_gauss} holds for product measures even 
under Euclidean distance~\cite{MT88}.
\end{itemize}

We now show how to approximate expectations of concentrated functions:
\begin{proposition}\label{th:dvc} Suppose that $F\colon\maty^n \to \mreals$ is $(b,c)$-concentrated
with respect to $P_{Y^n}^*$. Then
\begin{equation}\label{eq:dvc}
	 |\EE[F(Y^n)] - \EE[F(Y^{*n})]| \le 2\sqrt{c D(P_{Y^n} || P_{Y^n}^*)+ c\log b}\,,
\end{equation}
where
\begin{equation}  
	\EE[F(Y^{*n})] = \int F(y^n) dP_{Y^n}^*\,.\label{eq:expstar}
\end{equation}
\end{proposition}
\begin{IEEEproof}
Recall the Donsker-Varadhan inequality~\cite[Lemma
2.1]{DV75}: For any probability measures $P$ and $Q$ with $D(P||Q)<\infty$ and 
a measurable function $g$ such that $\int \exp\{g\} dQ < \infty$ we have that $\int g dP$ exists
(but perhaps is $-\infty$) and moreover
\begin{equation}\label{eq:dv}
	 \int g dP - \log \int \exp\{g\} dQ \le D(P||Q)\,.
\end{equation}

Since by~\eqref{eq:fcc} the moment generating function of $F$ exists under $P_{Y^n}^*$
, applying~\eqref{eq:dv} to $tF$ we get
\begin{equation}\label{eq:dv_ap}
	t \EE[ F(Y^n)] - \log \EE[\exp\{t F(Y^{*n})\}] \le D(P_{Y^n} || P_{Y^n}^*)\,.
\end{equation}
From~\eqref{eq:fcc} we have
\begin{equation}\label{eq:dv_aax}
	ct^2 - t \EE[ F(Y^n)] + t\EE[F(Y^{*n})] + D(P_{Y^n} || P_{Y^n}^*) + \log b \ge 0
\end{equation}
for all $t$. Thus the discriminant of the parabola in~\eqref{eq:dv_aax} must be non-positive which
is precisely~\eqref{eq:dvc}.
\end{IEEEproof}

Note that for empirical averages $F(y^n) = {1\over n} \sum_{j=1}^n f(y_i)$ we may either
apply the estimate for concentration in the example~\eqref{eq:ex2} and then use
Proposition~\ref{th:dvc}, or directly 
apply Proposition~\ref{th:dvc} to~\eqref{eq:dconv1}; the result is the same:
\begin{equation}
	\left| {1\over n}\sum_{j=1}^n \EE[ f(Y_j)] - \EE[f(Y^*)] \right| \le 2 \sqrt{{c\over n}
D(P_{Y^n} || P_{Y^n}^*)} \to 0\,,
\end{equation}
for any $f$ which is $(1,c)$-concentrated with respect to $P_Y^*$.

For the Gaussian channel, Proposition~\ref{th:dvc} and~\eqref{eq:conc_gauss}
yield:
\begin{corollary}\label{th:awgn2} For any $0<\epsilon<1$ there exist two constants $a_1, a_2>0$ such that for any
$(n, M, \epsilon)_{max,det}$ code for the $AWGN(P)$ channel and for any Lipschitz function 
$F\colon\mreals^n\to\mreals$ we have 
	\begin{equation}\label{eq:awgn2}
	 |\EE[F(Y^n)] - \EE[F(Y^{*n})]| \le a_1 \norm{F}_{Lip}\sqrt{nC - \log M_n + a_2 \sqrt{n}}\,, 
\end{equation}
where $C={1\over 2} \log (1+P)$ is the capacity.
\end{corollary}

Note that in the proof of Corollary~\ref{th:awgn2}, concentration of measure is used twice: once for 
$P_{Y^n|X^n}$ in the form of Poincar\'e's inequality (proof of Theorem~\ref{th:awgn1}) and
once in the form of~\eqref{eq:fcc} (proof of Proposition~\ref{th:dvc}).

\subsection{Concentration of functions of $Y^n$}\label{sec:approx_conc}

Not only can we estimate expectations of $F(Y^n)$ by
replacing the unwieldy $P_{Y^n}$ with the simple $P_{Y^n}^*$, but in fact 
the distribution of $F(Y^n)$ exhibits a sharp peak at its expectation:
\begin{proposition}\label{th:dv_conc} Consider a channel for which~\eqref{eq:re1} holds. Then for any $F$ which is
$(b,c)$-concentrated for such channel, we have for every $(n, M, \epsilon)_{max,det}$
code:
\begin{equation}\label{eq:re_conc1}
	 \PP[ |F(Y^n) - \EE[F(Y^{*n})]| > t] \le 3b \exp\left\{nC - \log M + a\sqrt{n} -
{t^2 \over 16 c}\right\} 
\end{equation}
and,
\begin{equation}\label{eq:re_var1}
	 \Var[F(Y^n)] \le 16 c \left(nC - \log M + a\sqrt{n} + \log(6be)\right)\,.
\end{equation}
\end{proposition}
\begin{IEEEproof} Denote for convenience:
\begin{eqnarray} \bar F &\eqdef& \EE[F(Y^{*n})]\,,\\
		\phi(x^n) &=& \EE[ F(Y^n) | X^n =x^n]\,.\label{eq:rec_0}
\end{eqnarray}
Then as a consequence of $F$ being $(b,c)$-concentrated for $P_{Y^n|X^n=x^n}$ we have
\begin{equation}\label{eq:rec_0b}
	 \PP[|F(Y^n) - \phi(x^n)| > t | X^n = x^n] \le b \exp\left\{-{t^2 \over
4c}\right\}\,. 
\end{equation}
Consider now a subcode $\matc_1$ consisting of all codewords such that $\phi(x^n) > \bar F + t$ for
$t>0$. The number $M_1 = |\matc_1|$ of codewords in this subcode is
\begin{equation}\label{eq:rec_0a}
	 M_1 = M\PP[\phi(X^n) > \bar F + t]\,.
\end{equation}
Let $Q_{Y^n}$ be the output distribution induced by $\matc_1$. We have the following
chain:
\begin{eqnarray}
 \bar F + t &\le& {1\over M_1} \sum_{x\in\matc_1} \phi(x^n) \label{eq:rec_1} \\
		&=& \int F(Y^n) dQ_{Y^n} \label{eq:rec_2} \\
		&\le& \bar F + 2 \sqrt{c D(Q_{Y^n} || P_{Y^n}^*) + c \log b} \label{eq:rec_3}\\
		&\le& \bar F + 2 \sqrt{c (nC - \log M_1 + a\sqrt{n}) + c\log b} \label{eq:rec_4}
\end{eqnarray}
where~\eqref{eq:rec_1} is by the definition of $\matc_1$,~\eqref{eq:rec_2} is
by~\eqref{eq:rec_0},~\eqref{eq:rec_3} is by Proposition~\ref{th:dvc} and the assumption of
$(b,c)$-concentration of $F$ under $P_{Y^n}^*$, and~\eqref{eq:rec_4} is by~\eqref{eq:re1}.

Together~\eqref{eq:rec_0a} and~\eqref{eq:rec_4} imply:
\begin{equation}\label{eq:rec_4a}
	 \PP[\phi(X^n) > \bar F + t] \le b \exp\left\{nC - \log M + a\sqrt{n} - {t^2\over
4c}\right\}\,.
\end{equation}
Applying the same argument to $-F$ we obtain a similar bound on $\PP[|\phi(X^n) - \bar{F} |>t ]$ and thus
\begin{eqnarray} 
\PP[|F(Y^n) - \bar F| > t] &\le& \PP[|F(Y^n) - \phi(X^n)| > t/2] + \PP[|\phi(X^n) -
\bar F| > t/2] \\
	&\le& b \exp\left\{-{t^2\over 16c}\right\} \left( 1 + 2 \exp\{nC -\log M +a\sqrt{n}\} \right)\label{eq:rec_5}\\
	&\le& 3b \exp\left\{-{t^2\over 16c} + nC -\log M +a\sqrt{n}\right\}
\label{eq:rec_6}\,,
\end{eqnarray}
where~\eqref{eq:rec_5} is by~\eqref{eq:rec_0b} and~\eqref{eq:rec_4a} and~\eqref{eq:rec_6}
is by~\eqref{eq:logm_up}.
Thus~\eqref{eq:re_conc1} is proven. Moreover,~\eqref{eq:re_var1} follows by~\eqref{eq:re_var}.
\end{IEEEproof}

Following up on Proposition \ref{th:dv_conc}, \cite{raginskyrefined} gives a bound, which in
contrast to  \eqref{eq:re_conc1}, shows explicit dependence on $\epsilon$.

\subsection{Relation to optimal transportation}\label{sec:opttrans}

Since the seminal work of Marton~\cite{KM86,KM96}, optimal transportation theory has emerged as one of the major tools for proving $(b,c)$-concentration 
of Lipschitz functions. Marton demonstrated that if a
probability measure $\mu$ on a metric space satisfies \textit{ a $T_1$ inequality }
\begin{equation}\label{eq:w1}
	 W_1(\nu, \mu) \le \sqrt{c' D(\nu||\mu)}\qquad \forall \nu
\end{equation}
then any Lipschitz $f$ is $(b, {\norm{f}_{Lip}^2 c})$-concentrated with respect to
$\mu$ for some
$b=b(c,c')$ and any $0<c<{c'\over 4}$. In~\eqref{eq:w1} $W_1(\nu, \mu)$ denotes the linear-cost
transportation distance, or Wasserstein-1 distance, defined as
\begin{equation}\label{eq:w1def}
	W_1(\nu, \mu) \eqdef \inf_{P_{YY'}} \EE[d(Y,Y')]\,,
\end{equation}
where $d(\cdot,\cdot)$ is the distance on the underlying metric space and the infimum is taken over
all couplings $P_{YY'}$ with fixed marginals $P_Y = \mu$, $P_{Y'}=\nu$. Note that
according to~\cite{dobrushin1970prescribing} we have $\norm{\nu - \mu}_{TV} = W_1(\nu, \mu)$
when the underlying distance on $\maty$ is $d(y, y') = 1\{y\neq y'\}$. 

In this section we show that~\eqref{eq:w1} in fact directly implies the estimate of
Proposition~\ref{th:dvc} without invoking either Marton's argument or Donsker-Varadhan inequality. Indeed, 
assume that $F\colon\maty^n\to\mreals$ is a Lipschitz function and observe that for any coupling
$P_{Y^n, Y^{*n}}$ we have
\begin{equation}\label{eq:ot_1}
	 |\EE[F(Y^n)] - \EE[F(Y^{*n})]| \le \norm{F}_{Lip} \EE[d(Y^n, Y^{*n})]\,,
\end{equation}
where the distance $d$ is either Hamming or Euclidean depending on the nature of $\maty^n$. Now taking the
infimum in the right-hand side of~\eqref{eq:ot_1} with respect to all couplings we observe
\begin{equation}
	|\EE[F(Y^n)] - \EE[F(Y^{*n})]| \le \norm{F}_{Lip} W_1(P_{Y^n}, P_{Y^n}*)
\end{equation}
and therefore by the transportation inequality~\eqref{eq:w1} we get
\begin{equation}
	|\EE[F(Y^n)] - \EE[F(Y^{*n})]| \le \sqrt{c' \norm{F}_{Lip}^2  D(P_{Y^n} || P_{Y^n}^*)}
\end{equation}
which is precisely what Proposition~\ref{th:dvc} yields for $(1, {c'\norm{F}_{Lip}^2\over 4})$-concentrated functions.

Our argument can be turned around and used to {\it prove} linear-cost transportation $T_1$
inequalities~\eqref{eq:w1}. Indeed, by the Kantorovich-Rubinstein duality~\cite[Chapter 1]{CV03} we have
\begin{equation}
	\sup_{F}|\EE[F(Y^n)] - \EE[F(Y^{*n})]| = W_1(P_{Y^n}, P_{Y^n}^*)\,,
\end{equation}
where the supremum is over all $F$ with $\norm{F}_{Lip} \le 1$. Thus the argument in the proof of
Proposition~\ref{th:dvc} shows that~\eqref{eq:w1} must hold for any $\mu$ for which every $1$-Lipschitz $F$ is
$(1,c')$-concentrated, demonstrating an equivalence between $T_1$
transportation and Gaussian-like concentration \textemdash  a result reported in~\cite[Theorem 3.1]{BG99lsi}.

We also mention that unlike general iid measures, an iid Gaussian $\mu =
\matn(0,1)^n$
satisfies a much stronger $T_2$-transportation inequality~\cite{MT96}
\begin{equation}\label{eq:w2}
	 W_2(\nu, \mu) \le \sqrt{c' D(\nu||\mu)}\qquad \forall \nu \ll \mu\,,
\end{equation}
where remarkably $c'$ does not depend on $n$ and the Wasserstein-2 distance
$W_2$ is defined as
\begin{equation}\label{eq:w2def}
	W_2(\nu, \mu) \eqdef \inf_{P_{YY'}} \sqrt{\EE[d^2(Y,Y')]}\,,
\end{equation}
the infimum being over all couplings as in~\eqref{eq:w1def}.

\subsection{Empirical averages of non-Lipschitz functions}\label{sec:approx_mean2}

One drawback of  relying on the transportation 
inequality~\eqref{eq:w1} in the proof of Proposition~\ref{th:dvc} is that
it does not show anything for non-Lipschitz functions. In this section we demonstrate how
the proof of Proposition~\ref{th:dvc} can be extended to functions that do not satisfy the strong
concentration assumptions.

\begin{proposition}\label{th:dvc_iid} Let $f\colon\maty\to\mreals$ be a (single-letter) function such that for
some $\theta > 0$ we have $m_1 \eqdef \EE[\exp\{\theta f(Y^*)\}] < \infty$
	(one-sided Cramer condition) and
$m_2 = \EE[f^2(Y^*)] < \infty$.
	Then there exists $b = b(m_1, m_2, \theta)>0$ such that for all $n\ge {16\over \theta^4}$ we have
\begin{equation}\label{eq:dvi}
	 {1\over n}\sum_{j=1}^n \EE[f(Y_j)] \le \EE[f(Y^*)] + \frac{1}{n^{3\over 4}}D(P_{Y^n} || P_{Y^n}^*) + \frac{b}{n^{\frac1{4}}} \end{equation}
\end{proposition}

\begin{IEEEproof} It is clear that if the moment-generating function $t\mapsto\EE[\exp\{t f(Y^*)\}]$ exists for
$t=\theta>0$ then it also exists for all $0\le t\le\theta$. Notice that since
\begin{equation}
	x^2 \exp\{-x\} \le 4 e^{-2} \log e\,,\qquad \forall x\ge0
\end{equation}
	we have for all $0 \le t \le {\theta \over 2}$:
\begin{eqnarray} \EE[f^2(Y^*) \exp\{t f(Y^*)\}] &\le& 
			\EE[f^2(Y^*) 1\{f < 0\} ] + {16 e^{-2} \log e\over (\theta - t)^2} 
				\EE\left[\exp\{\theta f(Y^*)\} 1\{f\ge 0\}\right]\\
			&\le& m_2 + {e^{-2} m_1 \log e\over (\theta - t)^2}\\
			&\le& m_2 + {4 e^{-2} m_1 \log e\over \theta^2}\\
			&\eqdef& b(m_1, m_2, \theta) \cdot 2\log e\,.\label{eq:dvi_1}
\end{eqnarray}
	Then a simple estimate
\begin{equation}\label{eq:dvi_2}
	 \log \EE[\exp\{t f(Y^*)\}] \le t \EE[f(Y^*)] + b t^2\,,\qquad 0\le t\le
{\theta\over2}\,,
\end{equation}
	can be obtained by taking the logarithm of the identity
\begin{equation}
	\EE[\exp\{t f(Y^*)\}] = 1 + {t\over \log e} \EE[f(Y^*)] + {1\over \log^2 e} \int_0^t ds
\int_0^s \EE[f^2(Y^*) \exp\{t f(Y^*)\}] du
\end{equation}
	and invoking~\eqref{eq:dvi_1} and $\log x \le {(x-1)\log e }$.

	Next, we define $ F(y^n) = {1\over n} \sum_{j=1}^n f(y_i) $
	and consider the chain:
\begin{eqnarray} 
	t \EE[F(Y^n)] &\le& \log \EE[\exp\{tF(Y^{*n})\}] + D(P_{Y^n} || P_{Y^n}^*)\label{eq:dvi_3} \\
			 &=& n  \log \EE[\exp\{{t\over n}f(Y^*)\}] + D(P_{Y^n} || P_{Y^n}^*)\label{eq:dvi_4} \\
			 &\le&  t \EE[f(Y^*)] + {b t^2\over n} + D(P_{Y^n} || P_{Y^n}^*)\,,\label{eq:dvi_5} 
\end{eqnarray}
where~\eqref{eq:dvi_3},~\eqref{eq:dvi_4},~\eqref{eq:dvi_5} follow from \eqref{eq:dv_ap}, $P_{Y^n}^* =
(P_Y^*)^n$ and~\eqref{eq:dvi_2} assuming $ {t\over n} \le {\theta \over 2}$.
	The proof concludes by letting $t={n^{3\over 4}}$ in~\eqref{eq:dvi_5}.
\end{IEEEproof}

A natural extension of Proposition~\ref{th:dvc_iid} to functions such as 
\begin{equation}
	F(y^n) = {1\over n-r+1} \sum_{j=1}^{n-r+1} f(y_j^{j+r-1})
\end{equation}
is made by replacing the step~\eqref{eq:dvi_4} with an estimate 
\begin{equation}\label{eq:dvi_remark}
	 \log \EE[ \exp\{tF(Y^*)\}] \le {n-r+1\over r} \log \EE\left[ \exp\left\{{tr\over
n} f(Y^{*r})\right\}\right]\,,
\end{equation}
which in turn is shown by splitting the sum into $r$ subsums with independent terms and then
applying Holder's inequality:
\begin{equation}
	\EE[ X_1 \cdots X_r ] \le \left( \EE[|X_1|^r]\cdots\EE[|X_1|^r]\right)^{1\over r}
\end{equation}

\subsection{Functions of degraded channel outputs}
Notice that if the same code is used over a channel $Q_{Y|X}$ which is stochastically
degraded with respect to $P_{Y|X}$ then by the data-processing for relative entropy, 
the upper bound~\eqref{eq:re1} holds for
$D(Q_{Y^n}||Q_{Y^n}^*)$, where $Q_{Y^n}$ is the output of the $Q_{Y|X}$ channel and $Q_{Y^n}^*$ is
the output of $Q_{Y|X}$ when the input is distributed according to a capacity-achieving
distribution of $P_{Y|X}$. Thus, in all the discussions the pair $(P_{Y^n},P_{Y^n}^*)$ can
be replaced with $(Q_{Y^n}, Q_{Y^n}^*)$ without any change in arguments or constants. This
observation can be useful in questions of information theoretic security, where the wiretapper has
access to a degraded copy of the channel output.

\subsection{Input distribution: DMC}\label{sec:dmcinp}

As shown in Section~\ref{sec:empir} we have for every $\epsilon$-capacity-achieving code:
\begin{equation}
	\bar P_n = {1\over n} \sum_{j=1}^n P_{Y_j} \to P_Y^*\,.
\end{equation}
As noted in~\cite{SV97}, convergence of output distributions can be propagated to 
statements about the input distributions.
This is obvious for the case of 
a DMC with a non-singular (more generally, injective) matrix $P_{Y|X}$. 
Even if the capacity-achieving input distribution is not unique, the following argument extends that of~\cite[Theorem 4]{SV97}. By
Theorem~\ref{th:ideq} and~\ref{th:dmc1} we know that
\begin{equation} \label{ahum}
	{1\over n} I(X^n; Y^n) \to C\,.
\end{equation}
Denote the single-letter empirical input distribution  by $P_{\bar X} = {1\over n} \sum_{j=1}^n P_{X_j}$.
Naturally, $ I(\bar X; \bar Y) \leq C $. However, in view of  \eqref{ahum}
and 
the concavity of mutual information, we must necessarily have
\begin{equation}
	I(\bar X; \bar Y) \to C\,,
\end{equation}
By compactness of the simplex of input
distributions and continuity of the mutual information on that simplex the distance to the (compact)
set of capacity achieving distributions $\Pi$ must vanish:
\begin{equation}\label{eq:di_cc}
	d(P_{\bar X}, \Pi) \to 0\,.
\end{equation}
If the capacity achieving distribution $P_X^*$ is unique, then~\eqref{eq:di_cc} shows the 
convergence of $P_{\bar X}\to P_X^*$ in the (strong) sense of total variation.

\subsection{Input distribution: AWGN}\label{sec:awgninp}

In the case of the AWGN, just like in the discrete case,~\eqref{eq:ideq} implies that 
for any capacity achieving sequence of codes we have
\begin{equation}\label{eq:awgn_inpconv}
	P_{\bar X}^{(n)} = {1\over n} \sum_{j=1}^n P_{X_j} \stackrel{w}{\to} P_X^* \eqdef \matn(0, P)\,,
\end{equation}
however, in the sense of weak convergence of distributions only. Indeed, 
the induced empirical output distributions satisfy 
\begin{equation}P_{\bar Y}^{(n)} =  P_{\bar X}^{(n)}* \matn(0,1)\,,\label{bossy}\end{equation}
where $*$ denotes convolution. By~\eqref{eq:ideq}, \eqref{bossy} converges in relative entropy
and thus weakly.
Consequently, characteristic functions of $P_{\bar Y}^{(n)}$ converge pointwise to that of
$\matn(0,1+P)$. By dividing out the
characteristic function of $\matn(0,1)$ (which is strictly positive), so do characteristic functions
of $P_{\bar X} ^{(n)}$. Then Levy's criterion establishes~\eqref{eq:awgn_inpconv}.

We now discuss whether~\eqref{eq:awgn_inpconv} can be claimed in a stronger topology than
the weak one. Since $P_{\bar X}$ is purely atomic and $P_X^*$ is purely diffuse, we have 
\begin{equation}
	||P_{\bar X} - P_X^*||_{TV} = 1\,,
\end{equation}
and convergence in total variation (let alone in relative entropy) cannot hold.

On the other hand, it is quite clear that the second
moment of ${1\over n} \sum P_{X_j}$ necessarily converges to that of $\matn(0,P)$. Together weak
convergence and control of second moments imply~\cite[(12), p.7]{CV03}
\begin{equation}\label{eq:ac_1}
	 W_2^2\left({1\over n} \sum_{j=1}^n P_{X_j}, P_X^*\right) \to 0\,.
\end{equation}
Therefore~\eqref{eq:awgn_inpconv} holds in the sense of topology metrized by the
$W_2$-distance.

Note that convexity properties of $W_2^2(\cdot, \cdot)$ imply
\begin{eqnarray}
	 W_2^2\left({1\over n} \sum_{j=1}^n P_{X_j}, P_X^*\right) &\le& {1\over n}
		\sum_{j=1}^n W_2^2\left(P_{X_j}, P_X^*\right) \\
		&\le& 	{1\over n} W_2^2\left(P_{X^n}, P_{X^n}^*\right)\,,\label{eq:ac_1a}
\end{eqnarray}
where we denoted
\begin{equation}
	P_{X^n}^* \eqdef (P_X^*)^n = \matn(0, P I_n)\,.
\end{equation}
Comparing~\eqref{eq:ac_1} and~\eqref{eq:ac_1a}, it is natural to conjecture
a stronger result: For any capacity-achieving  sequence of codes
\begin{equation}\label{eq:ac_conj}
	 {1\over \sqrt{n}} W_2(P_{X^n}, P_{X^n}^*) \to  0\,.
\end{equation}

Another reason to conjecture~\eqref{eq:ac_conj} arises from considering the
 behavior of Wasserstein distance under convolutions.
Indeed from the $T_2$-transportation inequality~\eqref{eq:w2} and the relative entropy bound~\eqref{eq:re1} we have 
\begin{equation}
	{1\over n} W^2_2(P_{X^n}*\matn(0, I_n), P_{X^n}^* * \matn(0,I_n)) \to 0\,,
\end{equation}
since by definition
\begin{eqnarray} 
	P_{Y^n} &=& P_{X^n} * \matn(0, I_n)\\
  	P_{Y^n}^* &=& P_{X^n}^* * \matn(0, I_n)\,,
\end{eqnarray}
where $*$ denotes convolution of distributions on $\mreals^n$.
Trivially, for any $P,Q$ and $\matn$ \textemdash  probability measures on $\mreals^n$ it is true
that (e.g.~\cite[Proposition 7.17]{CV03})
\begin{equation}\label{eq:ac_2}
	 W_2(P*\matn, Q*\matn) \le W_2(P,Q)\,. 
\end{equation}
Thus, overall we have
\begin{equation}
	0 \leftarrow {1\over \sqrt{n}} W_2(P_{X^n} * \matn(0, I_n), P_{X^n}^* *
\matn(0,I_n)) \le
{1\over\sqrt{n}} W_2(P_{X^n}, P_{X^n}^*)\,,
\end{equation}
and~\eqref{eq:ac_conj}implies that the convolution with the
Gaussian kernel is unable to significantly decrease $W_2$.

Despite the foregoing intuitive considerations,  conjecture~\eqref{eq:ac_conj} is false.
Indeed, define $D^*(M, n)$ to be the minimum achievable average
square distortion among all vector quantizers of the memoryless Gaussian source $\matn(0, P)$ for blocklength $n$
and cardinality $M$. In other words,
\begin{equation}
	D^*(M, n) = {1\over n} \inf_Q W_2^2(P_{X^n}^*, Q)\,,
\end{equation}
where the infimum is over all probability measures $Q$ supported on $M$ equiprobable atoms in
$\mreals^n$. The standard rate-distortion (converse) lower bound dictates
\begin{equation}
	{1\over n} \log M \ge {1\over 2} \log {P\over D^*(M,n)}
\end{equation}
and hence 
\begin{eqnarray} 
	 W_2^2(P_{X^n}, P_{X^n}^*) &\ge& nD^*(n, M)\\
				&\ge& nP \exp\left\{-{2\over n}\log M\right\}\,,
\end{eqnarray}
which shows that for any sequence of codes with $\log M_n = O(n)$, the normalized
transportation distance stays strictly bounded away from zero:
\begin{equation}
	\liminf_{n\to\infty} {1\over \sqrt{n}}W_2(P_{X^n}, P_{X^n}^*)  > 0\,.
\end{equation}

\medskip
\ifincludenorms
Nevertheless, assertion~\eqref{eq:awgn_inpconv} may be strengthened in several ways, see
Section~\ref{sec:awgn2}.
\fi

\subsection{Extension to other channels: tilting}\label{sec:eoc}

Let us review the scheme of investigating functions of the output $F(Y^n)$ 
that was employed in this paper so far.
First, an inequality~\eqref{eq:re1} was shown by verifying that $Q_Y = P_{Y^n}^*$ satisfies the
conditions of Theorem~\ref{th:sf}. Then an approximation of the form 
\begin{equation}\label{eq:approx2}
	F(Y^n) \approx \EE[F(Y^n)] \approx \EE[F(Y^{*n})]
\end{equation}
follows by Propositions~\ref{th:dvc} and~\ref{th:dv_conc} {\it simultaneously}
for all concentrated (e.g. Lipschitz) functions. In this way, all the channel-specific work is 
isolated in proving~\eqref{eq:re1}. On the other hand, verifying conditions of Theorem~\ref{th:sf}
for $Q_Y = P_{Y^n}^*$ may be quite challenging even for
memoryless channels. In this section we show how Theorem~\ref{th:sf} can be used 
to show~\eqref{eq:approx2} for a given
function $F$ in the absence of the universal estimate in~\eqref{eq:re1}.

Let $P_{Y|X}\colon \matx\to\maty$ be a random transformation,  $Y^\prime$ 
distributed according to auxiliary distribution $Q_Y$ and $F:\maty\to\mreals$ a function
such that
\begin{equation}\label{eq:tilt1}
	Z_F = \log \EE[\exp\{F(Y^{\prime})\}] < \infty\,,
\end{equation}
Let $Q^{(F)}_Y$ an $F$-tilting of $Q_Y$, namely
\begin{equation}\label{eq:tilt}
	 dQ^{(F)}_Y = \exp\{F - Z_F\} dQ_Y
\end{equation}

The core idea of our technique is that if $F$ is sufficiently regular and $Q_Y$ satisfies conditions of
Theorem~\ref{th:sf}, then $Q_Y^{(F)}$ also does.
Consequently, the expectation of $F$ under $P_Y$ (induced by the code) can be investigated in
terms of the moment-generating function of $F$ under $Q_Y$. For brevity we only
present a variance-based version (similar to Corollary~\ref{th:sfvar}):

\begin{theorem} Let $Q_Y$ and $F$ be such that~\eqref{eq:tilt1} holds and
\begin{eqnarray} 
		S &=& \sup_x \Var\left[\log {dP_{Y|X=x}\over Q_Y}(Y) \middle|X=x \right]
			< \infty\,,\\
		S_F &=& \sup_x \Var [F(Y)|X=x] \,.
\end{eqnarray}
	Then there exists a constant $a = a(\epsilon, S)>0$ such that 
	for any $(M, \epsilon)_{max,det}$ code we have for all $0\le t\le 1$
	\begin{eqnarray}\label{eq:theta_sf}
	 t \EE[F(Y)] - \log \EE[\exp\{t F(Y^\prime)\}] &\le& D(P_{Y|X} || Q_Y | P_X) - \log M +
a \sqrt{S + t^2 S_F}
	\end{eqnarray}
\end{theorem}
\begin{IEEEproof} Note that since
\begin{equation}
	\log {dP_{Y|X}\over dQ_Y^{(F)}} = \log {dP_{Y|X}\over dQ_Y} - F(Y) + Z_F
\end{equation}
	we have for any $0\le t\le 1$:
\begin{equation}
	D(P_{Y|X} || Q_Y^{(tF)}| P_X) = D(P_{Y|X} || Q_Y | P_X) - t \EE[F(Y)] + \log
\EE[\exp\{tF(Y^\prime)\}]\,,
\end{equation}
and from~\eqref{jensencauchy}
\begin{equation}
	\Var\left[ \log {dP_{Y|X}\over dQ_Y^{(F)}} \middle| X=x\right] \le 2 (S + t^2
S_F)\,,
\end{equation}
We conclude by invoking Corollary~\ref{th:sfvar}
with $Q_Y$ and $S$ replaced by $Q_Y^{(tF)}$ and $2S + 2t^2 S_F$, respectively.
\end{IEEEproof}

For example, Corollary~\ref{th:awgn2} is recovered from~\eqref{eq:theta_sf} by 
taking $Q_Y = P_{Y^n}^*$, applying~\eqref{eq:caod_prop}, estimating the moment-generating
function via~\eqref{eq:conc_gauss} and bounding $S_F$ via Poincar\'e inequality:
\begin{equation}\label{eq:poin1}
	 S_F \le b \norm{F}_{Lip}^2\,.
\end{equation}

\section{Binary hypothesis testing $P_{Y^n}$ vs. $P_{Y^n}^*$}\label{sec:stein}

We now turn to the question of distinguishing $P_{Y^n}$ from $P_{Y^n}^*$ in the sense of
binary hypothesis testing. First, a simple data-processing reasoning yields for any $0 < \alpha \leq 1$,
\begin{equation}\label{eq:dproc_beta}
	 d(\alpha || \beta_\alpha(P_{Y^n}, P_{Y^n}^*)) \le D(P_{Y^n} || P_{Y^n}^*)\,,
\end{equation}
where we have denoted the binary relative entropy
\begin{equation}
	d(x||y) \eqdef x\log {x\over y} + (1-x) \log {1-x \over 1-y}\,.
\end{equation}
From~\eqref{eq:re1} and~\eqref{eq:dproc_beta} we conclude: \textit{Every $(n, M,
\epsilon)_{max,det}$ code must satisfy}
\begin{equation}\label{eq:beta_weak}
	 \beta_\alpha(P_{Y^n}, P_{Y^n}^*) \ge \left( {M\over 2} \right)^{1\over \alpha}
		 \exp \left\{-n{C\over \alpha} - \sqrt{n}{a\over\alpha}\right\}
\end{equation}
for all  $0 < \alpha \le 1$.
Therefore, in particular we see that the hypothesis testing problem for discriminating
$P_{Y^n}$ from $P_{Y^n}^*$ has {\it zero Stein's exponent} $ - \frac1n \log \beta_\alpha(P_{Y^n}, P_{Y^n}^*) $, provided that the sequence of $(n, M_n,
\epsilon)_{max,det}$ codes  with output distribution $P_{Y^n}$, is capacity achieving.

The main result in this section gives a better bound than \eqref{eq:beta_weak}:
\begin{theorem}\label{th:stein} Consider one of the three types of channels introduced in
Section~\ref{sec:notation}. Then every $(n,
M, \epsilon)_{avg}$ code must satisfy
\begin{equation}\label{eq:beta_strong}
	 \beta_\alpha(P_{Y^n}, P_{Y^n}^*) \ge M \exp\{-nC - a_2 \sqrt{n}\} \qquad \epsilon
\le \alpha \le 1\,,
\end{equation}
where $a_2 = a_2(\epsilon, a_1)>0$ depends only on $\epsilon$ and the constant $a_1$ from~\eqref{eq:caod_var}.
\end{theorem}

To prove Theorem~\ref{th:stein} we introduce the following converse whose particular case $\alpha = 1$ is \cite[Theorem 27]{PPV08}:
\begin{theorem}\label{th:ss_meta}
	Consider an $(M, \epsilon)_{avg}$ code for an arbitrary random transformation
$P_{Y|X}$. Let $P_X$ be equiprobable on the codebook $\matc$ and $P_Y$ be the induced output distribution. Then for any $Q_Y$ and
$\epsilon \le \alpha \le 1$ we have
\begin{equation}\label{eq:ssavg}
	 \beta_{\alpha}(P_Y, Q_Y) \ge M \beta_{\alpha - \epsilon}(P_{XY}, P_X Q_Y)\,.
\end{equation}
	If the code is $(M, \epsilon)_{max,det}$ then additionally 
	\begin{equation}\label{eq:ssmax}
	 \beta_\alpha(P_Y, Q_Y) \ge {\delta \over 1-\alpha + \delta} 
		\,M \inf_{x\in\matc}\beta_{\alpha - \epsilon - \delta}(P_{Y|X=x}, Q_Y) \qquad
 \epsilon + \delta \le \alpha \le1\,,
\end{equation}
\end{theorem}
\begin{IEEEproof} For a given $(M, \epsilon)_{avg}$
code, define 
\begin{equation}
	Z = 1\{\hat W(Y) = W, Y\in E\}\,,
\end{equation}
where $W$ is the message and $E$ is an arbitrary event of the output space satisfying 
\begin{equation}\label{eq:ssavg1}
	 P_Y[E] \ge \alpha\,.
\end{equation}
As in the original meta-converse~\cite[Theorem 26]{PPV08} the main idea is to
use $Z$ as a suboptimal hypothesis test for discriminating $P_{XY}$ against $P_X Q_Y$.
Following the same reasoning as in~\cite[Theorem 27]{PPV08} one notices that
\begin{equation}
	(P_X Q_Y)[Z=1] \le {Q_Y[E] \over M}
\end{equation}
and
\begin{equation}
	P_{XY}[Z=1] \ge \alpha - \epsilon\,.
\end{equation}
Therefore, by definition of $\beta_\alpha$ we must have 
\begin{equation}\label{eq:ssavg2}
	 \beta_{\alpha-\epsilon}(P_{XY}, P_X Q_Y) \le {Q_Y[E] \over M}\,. 
\end{equation}
To complete the proof of~\eqref{eq:ssavg} we take the infimum in~\eqref{eq:ssavg2} over all $E$ satisfying~\eqref{eq:ssavg1}.
\par
To prove~\eqref{eq:ssmax}, we again consider any set $E$ satisfying~\eqref{eq:ssavg1}. Denote the codebook $\matc = \{ c_1 , \ldots , c_M \}$
and for $i = 1, \ldots, M$
\begin{eqnarray} p_i &=& P_{Y|X=c_i}[E]\\
   q_i &=& Q_Y[\hat W = i, E].
\end{eqnarray}
Since the sets $\{\hat W = i\}$ are disjoint, the (arithmetic) average of $q_i$ is
upper-bounded by
\begin{equation}
	\EE[q_W] \le {\frac1M }Q_Y[E]\,,
\end{equation}
whereas because of~\eqref{eq:ssavg1} we have
\begin{equation}
	\EE[p_W] \ge \alpha\,.
\end{equation}
Thus, the following lower bound holds:
\begin{equation}
	\EE\left[{Q_Y[E]\over M} p_W - \delta q_W\right] \ge  {Q_Y[E]\over M} (\alpha - \delta)
\end{equation}
implying that there must exist $i\in\{1,\ldots,M\}$ such that
\begin{equation}
	{Q_Y[E]\over M} p_i - \delta q_i \ge {Q_Y[E]\over M} (\alpha - \delta)\,.
\end{equation}
For such $i$ we clearly have
\begin{eqnarray} P_{Y|X=c_i}[E] &\ge& \alpha - \delta \\
	   Q_Y[\hat W = i, E] &\le& {Q_Y[E]\over M} {1-\alpha - \delta\over \delta}\,.
\end{eqnarray}
By the maximal probability of error constraint we deduce
\begin{equation}
	P_{Y|X=c_i}[E, \hat W =i] \ge \alpha - \epsilon - \delta
\end{equation}
and thus by the definition of $\beta_\alpha$:
\begin{equation}\label{eq:ssmax1}
	 \beta_{\alpha-\epsilon-\delta} (P_{Y|X=c_i}, Q_Y) \le {Q_Y[E]\over M} {1-\alpha
- \delta\over \delta} \,.
\end{equation}
Taking the infimum in~\eqref{eq:ssmax1} over all $E$ satisfying~\eqref{eq:ssavg1} completes the proof
of~\eqref{eq:ssmax}.
\end{IEEEproof}

\begin{IEEEproof}[Proof of Theorem~\ref{th:stein}]
To show~\eqref{eq:beta_strong} we first notice that as a consequence of~\eqref{eq:caod_prop},~\eqref{eq:caod_var} 
and~\cite[Lemma 59]{PPV08} (see also~\cite[(2.71)]{YP10}) we have for any $x^n\in\matx_n$:
\begin{equation}\label{eq:caod_beta}
	 \beta_\alpha(P_{Y^n | X^n =x^n}, P_{Y^n}^*) \ge {\alpha \over 2} \exp\left\{-nC -
\sqrt{2 a_1 n\over \alpha}\right\}\,.
\end{equation}
From~\cite[Lemma 32]{YP10} and the fact that the function of $\alpha$ in the right-hand
side of~\eqref{eq:caod_beta} is convex we obtain that for any $P_{X^n}$
\begin{equation}\label{eq:caod_beta2}
	 \beta_\alpha(P_{X^n Y^n}, P_{X^n} P_{Y^n}^*) \ge {\alpha \over 2} \exp\left\{-nC -
\sqrt{2 a_1 n\over \alpha}\right\}\,.
\end{equation}
Finally,~\eqref{eq:caod_beta2} and~\eqref{eq:ssavg} imply~\eqref{eq:beta_strong}.
\end{IEEEproof}

\section{AEP for the output process $Y^n$}\label{sec:aep}

Conventionally, we say that a sequence of distributions $P_{Y^n}$  on $\maty^n$ (with $\maty$ a countable
set) satisfies the asymptotic equipartition
property (AEP) if
\begin{equation}\label{eq:oaep}
	 {1\over n}\left|\log {1\over P_{Y^n}(Y^n)} - H(Y^n)\right| \to 0 
\end{equation}
in probability. In this section, we will take the AEP to mean convergence of \eqref{eq:oaep}
in the stronger sense of $L_2$, namely,
\begin{equation}\label{eq:oa_1}
	 \Var[\log P_{Y^n}(Y^n)] = o(n^2)\,, \qquad n\to\infty\,.
\end{equation}

\subsection{DMC}
Although the sequence of output distributions induced by a code is far from being 
(a finite chunk of) a stationary ergodic process, we will show that~\eqref{eq:oaep} is
satisfied for $\epsilon$-capacity-achieving codes (and other codes). Thus, in particular,
if the channel outputs are to be almost-losslessly compressed and stored for later decoding, ${1\over n}
H(Y^n)$ bits per sample would suffice (cf.~\eqref{eq:dmc1_entest}).
In fact, $\log {1\over P_{Y^n}(Y^n)}$ concentrates up to $\sqrt{n}$ around the
entropy $H(Y^n)$. Such questions are also interesting in other contexts and for other
types of distributions, see~\cite{VH97,BM11}.

\begin{theorem}\label{th:dmc_oaep2} Consider a DMC $P_{Y|X}$ with $C_1 < \infty$ (with or without
input constraints) and a capacity achieving
sequence of $(n, M_n, \epsilon)_{max,det}$ codes. Then the output AEP~\eqref{eq:oaep}
holds.
\end{theorem}
\begin{IEEEproof} 
In the proof of Theorem~\ref{th:dmc1} it was shown that $\log P_{Y^n}(y^n)$ is Lipschitz with
Lipschitz constant upper bounded by $a_1$. Thus by~\eqref{eq:conc_hamming}
and Proposition~\ref{th:dv_conc} we find that for any capacity-achieving sequence of codes
\eqref{eq:oa_1} holds.
\end{IEEEproof}

For many practically interesting DMCs (such as those with additive noise in
a finite group), the estimate~\eqref{eq:oa_1} can be improved
to $O(n)$ even without assuming the code to be capacity-achieving.

\begin{theorem}\label{th:dmc_oaep} Consider a DMC $P_{Y|X}$ with $C_1 < \infty$ (with or without
input constraints) and such that
$H(Y|X=x)$ is constant on $\matx$. Then for any sequence of $(n, M_n, \epsilon)_{max,det}$ codes
there exists a constant $a = a(\epsilon)$ such that for all $n$ sufficiently large
\begin{equation}
	\Var\left[ \log {P_{Y^n}(Y^n)} \right] \le a n\,.
\end{equation}
	In particular, the output AEP~\eqref{eq:oa_1} holds.
\end{theorem}
\begin{IEEEproof} First, let $X$ be a random variable and $A$ some event (think $\PP[A^c] \ll 1$)
such that 
\begin{equation}\label{eq:dmo_1} |X - \EE[X]| \le L
\end{equation}
if $X \not\in A$. 
Then, denoting $\Var[X|A] = \EE[X^2|A] - \EE^2[X|A]$,
\begin{eqnarray} \Var[X] &=& \EE[(X - \EE[X])^2 1_A] +\EE[(X - \EE[X])^2 1_{A^c}] \\
		   &\le&\EE[(X - \EE[X])^2 1_A] + \PP[A^c] L^2\label{eq:dmo_2}\\
		   &=& \PP[A] \left(\Var[X|A] + \left(\PP[A^c]\over \PP[A]\right)^2 
					(\EE[X] - \EE[X|A^c])^2 \right) + \PP[A^c] L^2 \label{eq:dmo_3}\\
		   &\le& \Var[X|A] + {\PP[A^c] \over \PP[A]} L^2\label{eq:dmo_4}\,, 
\end{eqnarray}
where~\eqref{eq:dmo_2} is by~\eqref{eq:dmo_1},~\eqref{eq:dmo_3} is because
\begin{eqnarray}
		\EE[(X - \EE[X])^2|A] &=& \Var[X|A] + (\EE[X|A] - \EE[X])^2 \\
&=& \Var[X|A] + \left({P[A^c]\over P[A]}\right)^2 (\EE[X] - \EE[X|A^c])^2 
\end{eqnarray}
which in turn follows from identity
\begin{equation}
   \EE[X|A] = {\EE[X] - \PP[A^c] \EE[X|A^c]\over \PP[A]}
\end{equation}
and~\eqref{eq:dmo_4} is because~\eqref{eq:dmo_1} implies $|\EE[X|A^c]
- \EE[X]|\le L$.
			 
Next, fix $n$ and for any codeword $x^n \in \matx_n$ denote for brevity
\begin{eqnarray} d(x^n) &=& D(P_{Y^n |X^n = x^n} || P_{Y^n})\\
	   v(x^n) &=& \EE\left[\log {1\over P_{Y^n}(Y^n)} \,\middle|\, X^n = x^n
\right]\\	
		 &=& d(x^n) + H(Y^n | X^n = x^n)\,.\label{eq:dmo_5b}
\end{eqnarray}
If we could show that for some $a_1>0$
\begin{eqnarray}\label{eq:dmo_5}
	 \Var[d(X^n)] \le a_1 n 
\end{eqnarray}
the proof would be completed as follows:
\begin{eqnarray} \Var\left[ \log {1\over P_{Y^n}(Y^n)} \right] &=&
		\Var\left[ \log {1\over P_{Y^n}(Y^n)} \,\middle|\, X^n \right] 	+ 
\Var[v(X^n)]\label{eq:dmo_6}\\
		&\le& a_2 n + \Var[v(X^n)]\label{eq:dmo_7}\\
		&=& a_2 n + \Var[d(X^n)]\label{eq:dmo_8}\\
		&\le& (a_1 + a_2)n\label{eq:dmo_9}\,,
\end{eqnarray}
where~\eqref{eq:dmo_7} follows for an appropriate constant $a_2>0$
from~\eqref{eq:dmc1_8},~\eqref{eq:dmo_8} is by~\eqref{eq:dmo_5b} and $H(Y^n|X^n=x^n)$ does
not depend on $x^n$ by assumption\footnote{This argument also shows how to construct a
counterexample when $H(Y|X=x)$ is non-constant: merge two constant composition subcodes
of types $P_1$ and $P_2$ such that $H(W|P_1) \neq H(W|P_2)$ where $W=P_{Y|X}$ is the
channel matrix. In this case one clearly has $\Var[\log P_{Y^n}(y^n)] \ge \Var[v(X^n)] =
\mathrm{const} \cdot n^2$.}, and~\eqref{eq:dmo_9} is by~\eqref{eq:dmo_5}.

To show~\eqref{eq:dmo_5}, first note the bound on the information density
\begin{equation}
	\imath_{X^n; Y^n}(x^n; y^n) = \log {P_{X^n|Y^n}(x^n | y^n) \over
P_{X^n}(x^n)} \le \log M_n\,.
\end{equation}
Second, as shown in~\eqref{eq:dmc1_7} one may take $S_m = a_3 n$ in Corollary~\ref{th:sfvar}. In
turn, this implies that one can take $\Delta = \sqrt{2 a_3 n\over 1-\epsilon}$ and $\delta' =
{1-\epsilon\over 2}$ in Theorem~\ref{th:sf}, that is:
\begin{equation}
	\inf_{x^n} \PP\left[ \log {P_{Y^n|X^n=x^n} \over P_{Y^n}}(Y^n) < d(x^n) + \Delta
\middle|X^n=x^n\right] \ge
{1+\epsilon\over 2}\,.
\end{equation}
Then applying Theorem~\ref{th:augustin} with $\rho(x^n) = d(x^n) + \Delta$ to the $(M_n', \epsilon)_{max,det}$ subcode
consisting of all codewords with $\{d(x^n) \le \log M_n - 2\Delta\}$ we get
\begin{equation}
	\PP[d(X^n) \le \log M_n - 2\Delta] \le {2\over 1-\epsilon} \exp\{-\Delta\}\,,
\end{equation}
since $M_n' = M_n \PP[d(X^n) \le \log M_n - 2\Delta]$ and
\begin{equation} 
	\EE[\exp(\rho(X^n)) | d(X^n) \le \log M_n - 2\Delta] \le  M_n \exp (- \Delta) \,.
\end{equation}
Now, we apply~\eqref{eq:dmo_4} to $d(X^n)$ with $L = \log M_n$ and $A = \{d(X^n) > \log M_n -
2\Delta\}$. Since $\Var[X|A] \le \Delta^2$ this yields
\begin{equation}\label{eq:dmo_10}
	 \Var[d(X^n)] \le \Delta^2 +{2 \log^2 M_n \over 1-\epsilon} \exp\{-\Delta\}  
\end{equation}
for all $n$ such that $ {2\over 1-\epsilon} \exp\{-\Delta\} \le {1\over 2} $. Since
$\Delta = O(\sqrt{n})$ and $\log M_n = O(n)$ we conclude from~\eqref{eq:dmo_10} that there
must be a constant $a_1$ such that~\eqref{eq:dmo_5} holds. 
\end{IEEEproof}
\subsection{AWGN}
Following the argument  of Theorem~\ref{th:dmc_oaep} step by step
with~\eqref{eq:awgn1_8} used in place of~\eqref{eq:dmc1_8}, we arrive at a
similar AEP for the AWGN channel.
\begin{theorem}\label{th:awgn_oaep} Consider the $AWGN$ channel. Then for any sequence of $(n, M_n, \epsilon)_{max,det}$ codes
there exists a constant $a = a(\epsilon)$ such that for all $n$ sufficiently large
\begin{equation}\label{eq:awgn_oaep}
	\Var\left[ \log { p_{Y^n}(Y^n)} \right] \le a n\,,
\end{equation}
	where $p_{Y^n}$ is the density of ${Y^n}$.
\end{theorem}

\begin{corollary} If in the setting of Theorem~\ref{th:awgn_oaep}, the codes are 
spherical (i.e., the energies of all codewords $X^n$ are equal) or, 
more generally,  
\begin{equation}\label{eq:awgn_oaepc2}
	\Var[||X^n||^2] = o(n^2),
\end{equation} then
\begin{equation}
	{1\over n} \left| \log {dP_{Y^n}\over dP_{Y^n}^*}(Y^n) - D(P_{Y^n}||P_{Y^n}^*)
\right | \to 0
\end{equation}
in $P_{Y^n}$-probability.
\end{corollary}
\begin{IEEEproof} To apply Chebyshev's inequality to $\log {dP_{Y^n}\over
dP_{Y^n}^*}(Y^n)$ we need, in addition to~\eqref{eq:awgn_oaep}, to show
\begin{equation}\label{eq:awgn_oaepc1}
	 \Var[\log p_{Y^n}^*(Y^n)] = o(n^2)\,, 
\end{equation}
	where $p_{Y^n}^*(y^n) = (2\pi (1+P))^{-{n\over 2}} e^{-{||y^n||^2\over 2 (1+P)}}$.
Introducing i.i.d. $Z_j \sim \matn(0,1)$ we have
	\begin{equation}\label{eq:awgn_oaepc3}
	 \Var[\log p_{Y^n}^*(Y^n)] = {\log^2 e \over 4(1+P)^2}\Var \left[||X^n||^2 + 2\sum_{j=1}^n
X_j Z_j + ||Z^n||^2\right]\,. 
\end{equation}
	The variances of the second and third terms are clearly $O(n)$, while the variance
of the first term is $o(n^2)$ by assumption~\eqref{eq:awgn_oaepc2}.
Then~\eqref{eq:awgn_oaepc3} implies~\eqref{eq:awgn_oaepc1} via \eqref{jensencauchy}.
\end{IEEEproof}

\ifincludenorms%
\section{Expectations of non-linear polynomials of Gaussian codes}\label{sec:awgn2}

This section contains results special to the AWGN channel. Because of the algebraic structure available on
$\mreals^n$ it is natural to ask whether we can provide approximations for polynomials. Since
Theorem~\ref{th:awgn1} shows the validity of~\eqref{eq:re1}, all the results for Lipschitz
(in particular linear) functions from Section~\ref{sec:relent} follow. Polynomials of
higher degree, however, do not admit bounded Lipschitz constants.
In this section we discuss the case of quadratic polynomials
(Section~\ref{sec:awgn_quad}) and  polynomials
of higher degree (Section~\ref{sec:lpnorm}). We present results directly in terms of the
polynomials in $(X_1,\ldots, X_n)$ on the 
\textit{input} space. This is strictly stronger than
considering polynomials on the output space, since $\EE[q(Y^n)] = \EE[q(X^n + Z^n)]$ and
thus by taking integrating over distribution of $Z^n$  problem reduces to computing
the expectation of a (different) polynomial of $X^n$. The reverse reduction is not
possible, clearly.

\subsection{Quadratic forms}\label{sec:awgn_quad}

We denote the canonical inner product on $\mreals^n$ as
\begin{equation}
	(\vect a, \vect b) = \sum_{j=1}^n a_j b_j\,,
\end{equation}
and write the quadratic form corresponding to matrix $\vect A$ as
\begin{equation}
	(\vect A \vect x, \vect x) = \sum_{j=1}^n \sum_{i=1}^n a_{i,j} x_i x_j\,.
\end{equation}
Note that when $X^n \sim \matn(0, P)^n$ we have trivially
\begin{equation}
	\EE[(\vect A X^n, X^n)] = P \tr \vect A\,,
\end{equation}
where $\tr$ is the trace operator. Therefore, the next result shows that the distribution
of good codes must be close to isotropic Gaussian distribution, at least in the sense of evaluating quadratic forms:

\begin{theorem}\label{th:qft} 
	For any $P>0$ and $0<\epsilon<1$ there exists a constant $b = b(P,\epsilon) >0$ such that
	for all $(n, M, \epsilon)_{max,det}$ codes and all quadratic forms $\vect A$ such that
\begin{equation}\label{eq:qft_0a}
	 - \vect I_n \le \vect A \le \vect I_n 
\end{equation}
	we have
\begin{equation}\label{eq:qft}
	|\EE[(\vect A X^n, X^n)] - P\tr \vect A| \le {2(1+P) \sqrt{n}\over \sqrt{\log e}} \sqrt{nC - \log M + b
\sqrt{n}}
\end{equation}
	and (a refinement for $\vect A=\vect I_n$)
\begin{equation}\label{eq:qft_a}
	 |\sum_{j=1}^n\EE[X_j^2] - nP| \le {2(1+P)\over \log e} 
			(nC - \log M + b \sqrt{n})\,.
\end{equation}
\end{theorem}
\begin{remark}
By using the same method as in the proof of Proposition~\ref{th:dv_conc} one can also
show that the estimate~\eqref{eq:qft} holds on a per-codeword basis for an overwhelming majority of
codewords.
\end{remark}

\begin{IEEEproof}
	Denote
\begin{eqnarray}
		 \bSigma &=& \EE [\vect x \vect x^T] \,,\\
		   \bV &=& (\bIn+\bSigma)^{-1}\,,\\
		   Q_{Y^n} &=& \matn(0, \bIn +\bSigma)\,,\\
		   R(\vect y|\vect x) &=& \log {dP_{Y^n|X^n=\vect x}\over dQ_{Y^n}}(\vect y)\,,\\
					&=& {\log e\over 2}\left( \ln \det (\bIn+\bSigma) + (\bV \vect y,\vect y)  -
			||\vect y - \vect x||^2 \right)\,,\\
		   d(\vect x) &=& \EE[R(Y^n|\vect x) | X^n=\vect x]\,,\\
			&=& {\log e\over 2} \left( \ln \det (\bIn+\bSigma) + (\bV\vect x,\vect x) + 
				\tr (\bV-\bIn) \right)\label{eq:qft_263} \\
		   v(\vect x) &=& \Var[R(Y^n|\vect x)|X^n = \vect x]\,.
\end{eqnarray}
Denote also the spectrum of $\bSigma$ by $\{\lambda_i, i=1,\ldots,n\}$ and its eigenvectors by 
$\{\vect v_i, i=1,\ldots,n\}$. We have then
\begin{eqnarray}	 
	\left| \EE[(\vect A X^n, X^n)] - P \tr \vect A\right| &=&  \left| \tr (\bSigma -P
\bIn) \vect A \right|\label{eq:qft_3a}\\
		&=& \left| \sum_{i=1}^n (\lambda_i - P)(\vect A \vect v_i, \vect v_i)
\right|\label{eq:qft_3b}\\
		&\le& \sum_{i=1}^n |\lambda_i - P|\,, \label{eq:qft_3}
\end{eqnarray}
where~\eqref{eq:qft_3b} follows by computing the trace in the eigenbasis of $\bSigma$ and~\eqref{eq:qft_3}
is by~\eqref{eq:qft_0a}.

From~\eqref{eq:qft_263}, it is straightforward to check that 
\begin{eqnarray} 
	D(P_{Y^n|X^n} || Q_{Y^n} | P_{X^n}) &=& \EE[d(X^n)]\\
		&=& {1\over 2} \log \det (\bIn+\bSigma)\\
		&=& {1\over 2} \sum_{j=1}^n \log (1 + \lambda_j)\,.\label{eq:qft_2}
\end{eqnarray}

	By using~\eqref{jensencauchy} we estimate
\begin{eqnarray}\label{eq:qft_1}
		v(\vect x) &\le& {3 \log^2 e} \left( {1\over 4} \Var[ ||Z^n||^2] + {1\over 4} \Var[(V
Z^n, Z^n)] + \Var[(V \vect x, Z^n)]\right)\\
		&\le& n \left({9\over 4} + 3P\right) \log^2 e \eqdef n b_1^2
\label{eq:qft_1x}
\end{eqnarray}
where~\eqref{eq:qft_1x} results from applying the following identities and bounds for $Z^n \sim \matn(0, I_n)$:
\begin{eqnarray} 
   	\Var[ ||Z^n||^2 ] &=& 3n\,,\\
	\Var[ (\vect a, Z^n) ] &=& ||\vect a||^2\,,\\
	\Var[ (\bV Z^n, Z^n) ] &=& 3 \tr \bV^2 \le 3n\\
	||\bV \vect x||^2 &\le& ||\vect x||^2 \le nP\,.
\end{eqnarray}
	Finally from Corollary~\ref{th:sfvar} applied with $S_m = b_1^2 n$ and~\eqref{eq:qft_2} we have
	\begin{eqnarray} 
		{1\over 2} \sum_{j=1}^n \log (1 + \lambda_j) &\ge& 
		\log M - b_1 \sqrt{n} -  \log {2 \over 1-\epsilon} \\
		&\ge& \log M - b \sqrt{n} \\
			&=& {n\over 2}(\log (1+P) - \delta_n)\,, \label{eq:qft_4}
	\end{eqnarray} 
	where we abbreviated
	\begin{eqnarray} 
		b &=& \sqrt{2 \left({9\over 4} + 3P\right) \over 1-\epsilon} \log e + \log {2 \over
1-\epsilon} \\
		\delta_n &=& 2(nC + b\sqrt{n} - \log M)\,.
\end{eqnarray}
	To derive~\eqref{eq:qft_a} consider the chain:
\begin{eqnarray} - \delta_n &\le& {1\over n} \sum_{j=1}^n \log{1+\lambda_i\over 1+P}\label{eq:qft_5}\\
		   &\le& \log \left({1\over n} \sum_{j=1}^n {1+\lambda_i\over 1+P}\right)\label{eq:qft_6}\\
		   &\le& {\log e\over n(1+P)} \sum_{j=1}^n (\lambda_i - P)\label{eq:qft_7}\\
		   &=& {\log e\over n(1+P)} (\EE[||X^n||^2] - nP)\label{eq:qft_8}
\end{eqnarray}
where~\eqref{eq:qft_5} is~\eqref{eq:qft_4},~\eqref{eq:qft_6} is by Jensen's
inequality,~\eqref{eq:qft_7} is by $\log x \le (x-1)\log e$. Note  that~\eqref{eq:qft_a} is equivalent to~\eqref{eq:qft_8}. 
Finally,~\eqref{eq:qft} follows from~\eqref{eq:qft_3},~\eqref{eq:qft_5} and the next Lemma
applied with $X$ equiprobable on $\left\{ {1+\lambda_i\over 1+P}, i=1,\ldots,n\right\}$.
\end{IEEEproof}
\begin{lemma}
	Let $X>0$ and $\EE[X] \le 1$, then
\begin{equation}\label{eq:pinsk1}
	 \EE[|X-1|] \le \sqrt{2\EE\left[\ln {1\over X}\right]} 
\end{equation}
\end{lemma}
\begin{IEEEproof} Define two distributions on $\mreals_+$:
\begin{eqnarray} P[E] &\eqdef& \PP[X \in E]\\
	   Q[E] &\eqdef& \EE[X \cdot 1\{X\in E\}] + (1-\EE[X]) 1\{0\in E\}\,.
\end{eqnarray}
Then, we have
\begin{eqnarray} 2||P-Q||_{TV} &=& 1-\EE[X] +  \EE[|X-1|]\\
		D(P||Q) &=& \EE\left[\log {1\over X}\right]\,. 
\end{eqnarray}
and~\eqref{eq:pinsk1} follows by~\eqref{eq:pinsker}.
\end{IEEEproof}

The proof of Theorem~\ref{th:qft} relied on a direct application of the main inequality (in the form of
Corollary~\ref{th:sfvar}) and is independent of the previous estimate~\eqref{eq:re1}. At the expense
of a more technical proof we could derive an order-optimal form of Theorem~\ref{th:qft} starting
from~\eqref{eq:re1} using concentration properties of Lipschitz functions. Indeed, 
notice that because $\EE[Z^n] = 0$ we have
\begin{equation}
	\EE[(\vect A Y^n, Y^n)] =  \EE[(\vect A X^n, X^n)] + \tr \vect A\,.
\end{equation}
Thus,~\eqref{eq:qft} follows from~\eqref{eq:re1} if we can show 
\begin{equation}\label{eq:qft_alt}
	 |\EE[(\vect A Y^n, Y^n)] - \EE[(\vect AY^{*n}, Y^{*n})]| \le b \sqrt{n D(P_{Y^n}||P_{Y^n}^*)}\,.
\end{equation}
This is precisely what Corollary~\ref{th:awgn2} would imply if the 
function $\vect y \mapsto (\vect A\vect y, \vect y)$ were Lipschitz with constant $O(\sqrt{n})$. 
However, $(\vect A\vect y, \vect y)$ is generally not Lipschitz when considered on the entire of
$\mreals^n$. On the other hand, it is clear that from the point of view of evaluation of both the 
$ \EE[(\vect A Y^n, Y^n)]$ and $\EE[(\vect A Y^{*n}, Y^{*n})]$ only vectors of norm $O(\sqrt{n})$ are important,
and when restricted to the ball $S=\{\vect y\colon \norm{\vect y}_2 \le b \sqrt{n}\}$
quadratic form $(\vect A \vect y, \vect y)$ does
have a required Lipschitz constant of $O(\sqrt{n})$. This approximation idea can be made precise 
using Kirzbraun's theorem (see~\cite{IJS53} for a short proof) to extend $(\vect A\vect y, \vect y)$ beyond the
ball $S$ preserving the maximum absolute value and the Lipschitz constant $O(\sqrt{n})$.
Another method of showing~\eqref{eq:qft_alt} is by using the Bobkov-G\"otze extension
of Gaussian concentration~\eqref{eq:conc_gauss} to non-Lipschitz functions~\cite[Theorem
1.2]{BG99lsi} to estimate the moment generating function of
$(\vect A Y^{*n}, Y^{*n})$ and apply~\eqref{eq:dv_ap} with $t = \sqrt{{1\over n}D(P_{Y^n}||P_{Y^n}^*)}$. Both methods
yield~\eqref{eq:qft_alt}, and hence~\eqref{eq:qft}, but with less sharp
constants than those in Theorem~\ref{th:qft}.

\subsection{Behavior of $||\vect x||_q$}\label{sec:lpnorm}

The next natural question is to consider polynomials of higher degree. The simplest example of such
polynomials are $F(\vect x)= \sum_{j=1}^n x_j^q$ for some power $q$, to analysis of which we proceed
now. 
To formalize the problem, consider $1\le q \le \infty$ and define the $q$-th norm of the input vector in the usual
way 
\begin{equation}
	||\vect x||_q \eqdef \left(\sum_{i=1}^n |x_i|^q\right)^{1\over q}\,.
\end{equation}

The aim of this section is to investigate the values of $||\vect x||_q$ for the codewords
of good codes for the AWGN channel. Notice that when the coordinates of $\vect x$ are independent Gaussians
we expect to have
\begin{equation}
	\sum_{i=1}^n |x_i|^q \approx n \EE[|Z|^q]\,,
\end{equation}
where $Z \sim \matn(0,1)$. In fact it can be shown that there exists a sequence of capacity achieving
codes and constants $B_q$, $1\le q\le \infty$ such that every codeword $\vect x$ at every
blocklength $n$ satisfies\footnote{This does not follow from a simple random coding argument since we want the
property to hold for every codeword, which constitutes exponentially many constraints.
However, the claim can indeed be shown by invoking the $\kappa\beta$-bound~\cite[Theorem
25]{PPV08} with a suitably
chosen constraint set $\sf F$.}:
\begin{equation}\label{eq:awl_0a}
	 ||\vect x||_q \le B_q n^{1\over q} = O(n^{1\over q}) \qquad 1 \le q < \infty\,,
\end{equation}
and
\begin{equation}\label{eq:awl_0b}
	 ||\vect x||_\infty \le B_{\infty} \sqrt{\log n} = O (\sqrt{\log n})\,.
\end{equation}
\textit{But do~\eqref{eq:awl_0a}-\eqref{eq:awl_0b} hold (possibly with different constants)
for {\em any} good code?} 

It turns out that the answer depends on the range of $q$ and
on the degree of optimality of the code. Our findings are summarized in Table~\ref{tab:lpnorm}.
The precise meaning of each entry will be clear from Theorems~\ref{th:lp_new1},
\ref{th:lp_new2} and their corollaries. The
main observation is that the closer the code size comes
to $M^*(n, \epsilon)$, the better $\ell_q$-norms reflect those of random Gaussian
codewords~\eqref{eq:awl_0a}-\eqref{eq:awl_0b}. Loosely speaking, very little can be said about
$\ell_q$-norms of capacity-achieving codes, while $O(\log n)$-achieving codes are almost
indistinguishable from the random Gaussian ones. In particular, we see that, for example, for
capacity-achieving codes it is not possible to approximate expectations of polynomials of degrees
higher than $2$ (or $4$ for dispersion-achieving codes) by assuming Gaussian inputs, since even the asymptotic growth rate with $n$ can be
dramatically different. The question of whether we can approximate expectations of arbitrary
polynomials for $O(\log n)$-achieving codes remains open.

\begin{table}[t]
\centering
\caption{Behavior of $\ell_q$ norms $\norm{\vect x}_q$ of codewords from codes for the AWGN
channel.}\label{tab:lpnorm}
\begin{tabular}{l||cccc}
Code & $1 \le q \le 2$ & $2 < q \le 4$ & $4 < q < \infty $ & $q=\infty$ \\[3pt]
\hline
\vrule width 0pt height 14pt 
random Gaussian & $n^{1\over q}$ &$n^{1\over q}$ &$n^{1\over q}$ & $\sqrt{\log n} $ \\
any $O(\log n)$-achieving & $n^{1\over q}$ &$n^{1\over q}$ &$n^{1\over q} \log^{q-4\over 2q} n$ & $\sqrt{\log n} $ \\
any dispersion-achieving & $n^{1\over q}$ &$n^{1\over q}$ & $o(n^{1\over 4})$ & $o(n^{1\over 4})$ \\
any $O(\sqrt{n})$-achieving & $n^{1\over q}$ &$n^{1\over q}$ & $n^{1\over 4}$ & $n^{1\over 4}$ \\
any capacity-achieving & $n^{1\over q}$ & $o(n^{1\over 2})$ & $o(n^{1\over 2})$ & $o(n^{1\over 2})$ \\
any code & $n^{1\over q}$ & $n^{1\over 2}$ &$n^{1\over 2}$ &$n^{1\over 2}$ \\
\hline
\end{tabular}
\vskip 3pt
Note: All estimates, except $n^{1\over q} \log^{q-4\over 2q} n$, are shown to be tight.~\hskip 50pt{}
\end{table}

We proceed to support the statements made in Table~\ref{tab:lpnorm}. 

In fact, each estimate in Table~\ref{tab:lpnorm}, except $n^{1\over q} \log^{q-4\over 2q}
n$, is tight in the following sense: if the entry is $n^\alpha$, then there exists a constant $B_q$
and a sequence of $O(\log n)$-, dispersion-, $O(\sqrt{n})$-, or capacity-achieving $(n, M_n,
\epsilon)_{max,det}$ codes such that \textit{each} codeword $\vect x \in \mreals^n$ satisfies
for all $n\ge 1$
\begin{equation}\label{eq:lp_low1}
	\norm{\vect x}_q \ge B_q n^\alpha\,. 
\end{equation}
If the entry in the table states $o(n^\alpha)$ then there is $B_q$ such that 
for \textit{any} sequence $\tau_n\to0$ there exists a
sequence of $O(\log n)$-, dispersion-, $O(\sqrt{n})$-, or capacity-achieving $(n, M_n,
\epsilon)_{max,det}$ codes such that each codeword satisfies for all $n\ge 1$
\begin{equation}\label{eq:lp_low2}
	\norm{\vect x}_q \ge B_q \tau_n n^\alpha\,. 
\end{equation}

First, notice that a code from any row is an example of a code for the next row, so we only need to
consider each entry which is worse than the one directly above it. Thus it suffices to show the tightness
of $o(n^{1\over 4})$, $n^{1\over 4}$, $o(n^{1\over 2})$ and $n^{1\over 2}$.

To that end recall that by~\cite[Theorem 54]{PPV08} the maximum 
number of codewords $M^*(n, \epsilon)$ at a fixed probability of error $\epsilon$ for the
AWGN channel satisfies 
\begin{equation}\label{eq:awl_0c}
	 \log M^*(n, \epsilon) = n C - \sqrt{n V} Q^{-1}(\epsilon) + O(\log n)\,,
\end{equation}
where $V(P) = {\log^2 e\over 2} {P(P+2)\over (P+1)^2}$ is the channel dispersion.
Next, we fix a sequence $\delta_n \to 0$, such that $n\delta_n \to \infty$ and construct the following sequence of
codes. The first coordinate $x_1 = \sqrt{n \delta_n P}$ for every codeword and the rest
$(x_2, \ldots, x_n)$ are chosen as coordinates of an optimal AWGN code for blocklength
$n-1$ and power-constraint $(1 - \delta_n) P$. Following the argument of~\cite[Theorem
67]{PPV08} the number of codewords $M_n$ in such a code will be at least 
\begin{align}
	 \log M_n &= (n-1) C(P - \delta_n) - \sqrt{(n-1)V(P-\delta_n)} Q^{-1}(\epsilon) + O(1)\\
		  &= nC(P) - \sqrt{nV(P)} Q^{-1}(\epsilon) + O(n\delta_n)\label{eq:awl_3}\,.
\end{align}
At the same time, because $x_1$ of each codeword $\vect x$ is abnormally high we have
\begin{equation}\label{eq:awl_4}
	 \norm{\vect x}_q \ge \sqrt{n \delta_n P}\,. 
\end{equation}
So all the examples are constructed by choosing a suitable $\delta_n$ as follows:
\begin{itemize}
\item Row 1: see~\eqref{eq:awl_0a}-\eqref{eq:awl_0b}.
\item Row 2: nothing to prove.
\item Row 3: for entries $o(n^{1\over 4})$ taking $\delta_n = {\tau_n^2\over \sqrt{n}}$ yields a
dispersion-achieving code according to~\eqref{eq:awl_3}; the estimate~\eqref{eq:lp_low2} follows
from~\eqref{eq:awl_4}.
\item Row 4: for entries $n^{1\over 4}$  taking $\delta_n = {1\over \sqrt{n}}$ yields an
$O(\sqrt{n})$-achieving code according to~\eqref{eq:awl_3}; the estimate~\eqref{eq:lp_low1} follows
from~\eqref{eq:awl_4}.
\item Row 5: for entries $o(n^{1\over 2})$ taking $\delta_n = \tau_n^2$ yields a
capacity-achieving code according to~\eqref{eq:awl_3}; the estimate~\eqref{eq:lp_low2} follows
from~\eqref{eq:awl_4}.
\item Row 6: for entries $n^{1\over 2}$ we can take a codebook with one codeword $(\sqrt{nP}, 0,
\ldots, 0)$.
\end{itemize}
\begin{remark}
 The proof can be modified to show that in each case there are codes that 
simultaneously achieve all entries in the respective row of
Table~\ref{tab:lpnorm} (except $n^{1\over q} \log^{q-4\over 2q} n$).
\end{remark}

We proceed to proving upper bounds.
First, we recall some simple relations between the $\ell_q$ norms of vectors in $\mreals^n$. To
estimate a lower-$q$ norm in terms of a higher one, we invoke Holder's inequality:
\begin{equation}\label{eq:al_qlow}
	 \norm{\vect x}_q \le n^{{1\over q} - {1\over p}} \norm{\vect x}_p\,,\qquad 1\le q
\le p \le \infty\,.
\end{equation}
To provide estimates for $q>p$, notice that obviously
\begin{equation}
	\norm{\vect x}_\infty \le \norm{\vect x}_p\,.
\end{equation}
Then, we can extend to $q<\infty$ via the following chain:
\begin{eqnarray} 
	\norm{\vect x}_q &\le& \norm{\vect x}_\infty^{1-{p\over q}} \norm{\vect x}_p^{p\over q}
			\label{eq:al_qhigh1}\\
		   &\le& \norm{\vect x}_p\,,\qquad  q \ge p \label{eq:al_qhigh2}
\end{eqnarray}
Trivially, for $q=2$ the answer is given by the power constraint
\begin{equation}\label{eq:awl_1}
	 \norm{\vect x}_2 \le \sqrt{nP} 
\end{equation}
Thus by~\eqref{eq:al_qlow} and~\eqref{eq:al_qhigh2} we get: \textit{Each codeword of any code for the
$AWGN(P)$ channel must satisfy}
\begin{equation}\label{eq:awl_2}
	 \norm{\vect x}_q \le \sqrt{P} \cdot \begin{cases} n^{1\over q},& 1\le q\le 2\,,\\
					    n^{1\over 2},& 2 < q\le \infty\,.
				\end{cases}
\end{equation}
This proves the entries in the first column and the last row of Table~\ref{tab:lpnorm}.

Before proceeding to justify the upper bounds for $q>2$ we point out an obvious problem with trying to estimate
$\norm{\vect x}_q$ for \textit{each} codeword.
Given any code whose codewords lie exactly on the power sphere, 
we can always apply an orthogonal transformation to it so that one of the
codewords becomes $(\sqrt{nP}, 0, 0, \ldots 0)$. For such a codeword we have
\begin{equation}
	\norm{\vect x}_q = \sqrt{nP}
\end{equation}
and the upper-bound~\eqref{eq:awl_2} is tight. Therefore, to improve upon the~\eqref{eq:awl_2} we
must necessarily consider subsets of codewords of a given code. For simplicity below we show
estimates for the \textit{half} of all codewords.

The following result, proven in the Appendix, takes care of the sup-norm:
\begin{theorem}[$q=\infty$]\label{th:lp_new1} For any $0<\epsilon < 1$ and $P>0$ there exists a
constant $b=b(P,\epsilon)$ such that for any%
\footnote{$N(P,\epsilon) = 8(1+2P^{-1}) (Q^{-1}(\epsilon))^2$ for $\epsilon <
{1\over2}$ and $N(P,\epsilon)=1$ for $\epsilon \ge {1\over2}$.} %
$n\ge N(P,\epsilon)$ and any $(n, M, \epsilon)_{max,det}$-code for the $AWGN(P)$ channel 
at least \textit{half} of the codewords satisfy
\begin{equation}\label{eq:lpn1}
	 \norm{\vect{x}}_\infty^2 \le {4(1+P)\over \log e} \left(nC - \sqrt{nV} Q^{-1}(\epsilon) + 2
\log n + \log b - \log {M\over 2}\right)\,,
\end{equation}
where $C$ and $V$ are the capacity and the dispersion. In particular, the expression in
$(\cdot)$ is non-negative for all codes and blocklengths.
\end{theorem}

\begin{remark}
 What sets Theorem~\ref{th:lp_new1} apart from other results is its sensitivity to 
whether the code achieves the dispersion term. This is unlike estimates of the form~\eqref{eq:re1}, 
which only sense whether the code is $O(\sqrt{n})$-achieving or not.
\end{remark}

From Theorem~\ref{th:lp_new1} the explanation of the entries in the last column of
Table~\ref{tab:lpnorm} becomes obvious: the more terms the code achieves in the asymptotic expansion
of $\log M^*(n, \epsilon)$ the closer $\norm{\vect x}_\infty$ becomes to the
$O(\sqrt{\log n})$, which arises from a random Gaussian codeword~\eqref{eq:awl_0b}.
To be specific, we give the following exact statements:
\begin{corollary}[$q=\infty$ for $O(\log n)$-codes]\label{th:lp_up1} For any $0<\epsilon < 1$ and $P>0$ there exists a constant
$b=b(P,\epsilon)$ such that for any $(n, M_n, \epsilon)_{max,det}$-code for the $AWGN(P)$ with
\begin{equation}\label{eq:lup1_a}
	 \log M_n \ge nC - \sqrt{nV} Q^{-1}(\epsilon) - K \log n 
\end{equation}
for some $K>0$ we have that at least \textit{half} of the codewords
satisfy
\begin{equation}\label{eq:lup1_b}
	 \norm{\vect{x}}_\infty \le \sqrt{(b+K)\log n}+b\,.
\end{equation}
\end{corollary}
\begin{corollary}[$q=\infty$ for capacity-achieving codes]\label{th:lp_up2} 
For any capacity-achieving sequence of $(n, M_n, \epsilon)_{max,det}$-codes there exists a sequence
$\tau_n\to0$ such that for at least \textit{half} of the codewords we have
\begin{equation}\label{eq:lup2_a}
	 \norm{\vect{x}}_\infty \le \tau_n n^{1\over 2}\,.
\end{equation}
Similarly, for any dispersion-achieving sequence of $(n, M_n, \epsilon)_{max,det}$-codes there exists a sequence
$\tau_n\to0$ such that for at least \textit{half} of the codewords we have
\begin{equation}\label{eq:lup2_b}
	 \norm{\vect{x}}_\infty \le \tau_n n^{1\over 4}\,.
\end{equation}
\end{corollary}

\begin{remark}
By~\eqref{eq:lp_low2}, the sequences $\tau_n$ in Corollary~\ref{th:lp_up2} are necessarily code-dependent.
\end{remark}

For the $q=4$ we have the following estimate (see Appendix for the proof):
\begin{theorem}[$q=4$]\label{th:lp_new2} For any $0<\epsilon < {1\over2}$ and $P>0$ there exist
constants $b_1>0$ and $b_2>0$, depending on $P$ and $\epsilon$, such that for
 any $(n, M, \epsilon)_{max,det}$-code for the $AWGN(P)$ channel 
at least \textit{half} of the codewords satisfy
\begin{equation}\label{eq:lpn2}
	\norm{\vect x}_4^2 \le {2\over b_1} \left(nC + b_2 \sqrt{n} - \log {M\over 2}\right)\,,
\end{equation}
where $C$ is the capacity of the channel. In fact, we also have a lower bound
\begin{equation}\label{eq:awgn4}
	 \EE[ \norm{\vect x}_4^4 ] \ge 3n P^2 - (nC - \log M + b_3\sqrt{n}) n^{1\over 4}\,, 
\end{equation}
for some $b_3 = b_3(P, \epsilon)>0$.
\end{theorem}
\begin{remark}
Note that $\EE[ \norm{\vect z}_4^4] = 3n P^2$ for $\vect z \sim \matn(0, P)^n$.
\end{remark}

We can now complete the proof of the results in Table~\ref{tab:lpnorm}:
\begin{enumerate}
\item Row 2: $q=4$ is Theorem~\ref{th:lp_new2};~$2 < q\le 4$ follows by~\eqref{eq:al_qlow} with $p=4$;~
$q=\infty$ is Corollary~\ref{th:lp_up1}; for $4<q<\infty$ we apply interpolation via~\eqref{eq:al_qhigh1} with $p=4$.
\item Row 3: $q\le4$ is treated as in Row 2; $q=\infty$ is Corollary~\ref{th:lp_up2}; for $4<q<
\infty$ apply interpolation~\eqref{eq:al_qhigh1} with $p=4$.
\item Row 4: $q\le 4$ is treated as in Row 2;~$q\ge 4$ follows from~\eqref{eq:al_qhigh2} with $p=4$.
\item Row 5: $q=\infty$ is Theorem~\ref{th:lp_up2}; for $2<q<\infty$ we apply
interpolation~\eqref{eq:al_qhigh1} with $p=2$.
\end{enumerate}

The upshot of this section is that we cannot approximate values of non-quadratic polynomials in $\vect x$ (or
$\vect y$) by assuming iid Gaussian entries, unless the code is $O(\sqrt{n})$-achieving, in which case we can go up to
degree $4$ but still will have to be content with one-sided (lower) bounds only, cf.~\eqref{eq:awgn4}.\footnote{Using quite 
similar methods,~\eqref{eq:awgn4} can be extended to certain bi-quadratic
forms, i.e. 4-th degree polynomials $\sum_{i,j} a_{i-j} x_i^2 x_j^2$, where $A=(a_{i-j})$ is a Toeplitz positive semi-definite matrix.}

Before closing this discussion we demonstrate the sharpness of the arguments in this section
 by considering the following example. Suppose that a power of a 
codeword $\vect x$ from a capacity-dispersion optimal code is measured by an imperfect
tool, such that its reading is described by
\begin{equation}
	\mathcal{E} = {1\over n} \sum_{i=1}^n (x_i)^2 H_i\,,
\end{equation}
where $H_i$'s are i.i.d bounded random variables with expectation and variance both equal
to $1$. For large blocklengths $n$ we expect $\mathcal{E}$ to be Gaussian with mean $P$ and
variance ${1\over n}\norm{\vect x}_4^4$. On the one hand, Theorem~\ref{th:lp_new2} shows
that the variance will not explode;~\eqref{eq:awgn4} shows that it will be at least
as large as that of a Gaussian codebook. Finally, to establish the asymptotic normality
rigorously, the usual approach based on checking Lyapunov condition will fail as shown
by~\eqref{eq:lp_low2}, but the Lindenberg condition does hold as a consequence of
Theorem~\ref{th:lp_up2}. If in addition, the code is $O(\log n)$-achieving then
\begin{equation}
	\PP[|\mate - \EE[\mate]| > \delta] \le e^{-{n \delta^2\over b_1 + b_2 \delta \sqrt{\log n}}}\,.
\end{equation}

\appendix

In this appendix we prove results from Section~\ref{sec:lpnorm}.

To prove Theorem~\ref{th:lp_new1} the basic intuition is that any codeword which is abnormally peaky (i.e., has a high value of 
$\norm{\vect{x}}_\infty$) is wasteful in terms of allocating its power budget. Thus a good
 capacity- or dispersion-achieving codebook cannot have too many of such wasteful codewords.
 A non-asymptotic formalization of this intuitive argument is as follows:

\begin{lemma}\label{th:linf} For any $\epsilon \le {1\over 2}$ and $P>0$ there exists a constant $b=b(P,\epsilon)$
such that given any $(n, M, \epsilon)_{max,det}$ code for the $AWGN(P)$ channel, we have for any
$0\le \lambda \le P$:\footnote{For $\epsilon > {1\over2}$ one must replace $V(P-\lambda)$ with $V(P)$
in~\eqref{eq:linf}. This does not modify any of the arguments required to prove
Theorem~\ref{th:lp_new1}.}
\begin{equation}\label{eq:linf}
	 \PP[\norm{X^n}_\infty \ge \sqrt{\lambda n}] \le {b\over M} 
		\exp\left\{ nC(P-\lambda) - \sqrt{nV(P-\lambda)} Q^{-1}(\epsilon) + 2\log n
\right\}\,,
\end{equation}
where $C(P)$ and $V(P)$ are the capacity and the dispersion of the $AWGN(P)$ channel, and $X^n$ is
the output of the encoder assuming equiprobable messages.
\end{lemma}
\begin{IEEEproof}
Our method is to apply the meta-converse in the form of~\cite[Theorem 30]{PPV08} to the subcode that
satisfies $\norm{X^n}_\infty \ge \sqrt{\lambda n}$. Application of a meta-converse requires selecting
a suitable auxiliary channel $Q_{Y^n|X^n}$. We specify this channel now. For any $\vect x\in\mreals^n$ 
let $j^*(\vect x)$ be the first index s.t. $|x_j| = ||\vect x||_\infty$, then we set 
\begin{equation}\label{eq:linf_q}
	Q_{Y^n|X^n}(y^n | \vect x) = P_{Y|X}(y_{j^*}|x_{j^*}) \prod_{j\neq j^*(\vect x)} P_Y^*(y_j)
\end{equation}
We will show below that for some $b_1 = b_1(P)$ any $M$-code over the $Q$-channel~\eqref{eq:linf_q} 
has average probability of error $\epsilon'$ satisfying:
\begin{equation}\label{eq:li_0}
	 1-\epsilon' \le {b_1 n^{3\over2}\over M}\,.
\end{equation}
On the other hand, writing the expression for  $\log {dP_{Y^n|X^n=\vect x} \over dQ_{Y^n|X^n=\vect
x}}(Y^n)$ we see that it coincides with the expression for $\log {dP_{Y^n|X^n=\vect x} \over
dP_{Y^n}^*}$ except that the term corresponding to $j^*(\vect x)$ will be missing; compare with~\cite[(4.29)]{YP10}. Thus, one can
repeat step by step the analysis in the proof of~\cite[Theorem 65]{PPV08} with the only difference
that $nP$ should be replaced by $nP - \norm{\vect x}_{\infty}^2$ reflecting the reduction in the
energy due to skipping of $j^*$. Then, we obtain for some $b_2 = b_2(\alpha, P)$:
\begin{equation}\label{eq:li_1}
	 \log \beta_{1-\epsilon}(P_{Y^n |X^n = \vect x}, Q_{Y^n | X^n = \vect x}) \ge 
 -n C\left(P-{\norm{\vect x}_\infty^2\over n}\right) + \sqrt{n V\left(P-{\norm{\vect x}_\infty^2\over n}\right)}
Q^{-1}(\epsilon) - {1\over 2} \log n - b_2\,,
\end{equation}
which holds simultaneously for all $\vect x$ with $\norm{\vect x} \le \sqrt{nP}$. Two remarks are
in order: first, the analysis in~\cite[Theorem 64]{PPV08} must be done replacing $n$ with $n-1$, but
this difference is absorbed into $b_2$. Second, to see that $b_2$ can be chosen independent of $\vect x$ notice that $B(P)$ in~\cite[(620)]{PPV08} tends to
$0$ with $P\to0$ and hence can be bounded uniformly for all $P\in[0, P_{max}]$.

Denote the cardinality of the subcode $\{\norm{\vect x}_\infty \ge \sqrt{\lambda
n}\}$ by
\begin{equation}
	M_\lambda = M \PP[\norm{\vect x}_\infty \ge \sqrt{\lambda n}]\,.
\end{equation}
Then according to~\cite[Theorem 30]{PPV08}, we get
\begin{equation}
	\inf_{\vect x} \beta_{1-\epsilon}(P_{Y^n |X^n = \vect x}, Q_{Y^n | X^n = \vect x}) \le
1-\epsilon'\,,
\end{equation}
where the infimum is over the codewords of $M_\lambda$-subcode. Applying both~\eqref{eq:li_0}
and~\eqref{eq:li_1} we get
\begin{equation}
	\inf_{\vect x} \left( -n C\left(P-{\norm{\vect x}_\infty^2\over n}\right) + \sqrt{n V\left(P-{\norm{\vect x}_\infty^2\over n}\right)}
Q^{-1}(\epsilon)\right) - {1\over 2} \log n - b_2  \le -\log M_\lambda + \log b_1 + {3\over 2} \log
n\label{eq:linf_1}
\end{equation}
and, further, since the function of $||\vect x||_\infty$ in left-hand side of~\eqref{eq:linf_1} is monotone in $\norm{\vect x}_\infty$:
\begin{equation}
	-n C(P-\lambda) + \sqrt{n V(P-\lambda)}
Q^{-1}(\epsilon) - {1\over 2} \log n - b_2  \le -\log M_\lambda + \log b_1 + {3\over 2} \log
n\,.
\end{equation}
Thus, overall
\begin{equation}
	\log M_\lambda \le n C(P-\lambda) - \sqrt{n V(P-\lambda)}
Q^{-1}(\epsilon) + 2\log n + b_2 + \log b_1\,,
\end{equation}
which is equivalent to~\eqref{eq:linf} with $b=b_1\exp\{b_2\}$.

It remains to show~\eqref{eq:li_0}. Consider an $(n, M, \epsilon')_{avg,det}$-code for the 
$Q$-channel and let $M_j, j=1,\ldots,n$ denote the cardinality of the set of all codewords with
$j^*(\vect x) = j$. Let $\epsilon'_j$ denote the minimum possible average probability of error of
each such codebook achievable with the maximum likelihood (ML) decoder.
Since
\begin{equation}
	1-\epsilon' \le {1\over M} \sum_{j=1}^n M_j (1-\epsilon'_j)
\end{equation}
it suffices to prove
\begin{equation}\label{eq:li_f}
	 1-\epsilon'_j \le {\sqrt{2nP\over \pi} + 2 \over M_j} 
\end{equation}
for all $j$. 
Without loss of generality assume $j=1$ in which case the observations $Y_2^n$ are useless for determining
the value of the true codeword. Moreover, the ML decoding regions $D_i, i=1,\ldots, M_j$ for each
codeword are disjoint intervals in $\mreals^1$ (so that decoder outputs message estimate $i$
whenever $Y_1\in D_i$). Note that for $M_j \le 2$ there is nothing to prove, so assume otherwise. 
Denote the first coordinates of the $M_j$ codewords by $x_i, i =1,\ldots, M_j$ and assume
(without loss of generality) that $-\sqrt{nP} \le x_1 \le x_2 \le \cdots \le x_{M_j} \le \sqrt{nP}$ 
and that $D_2,\ldots
D_{M_j-1}$ are finite intervals. We have the following chain then
\begin{eqnarray} 
	1-\epsilon'_j &=& {1\over M_j} \sum_{i=1}^{M_j} P_{Y|X}(D_i|x_i)\label{eq:li_2}\\
		 &\le&	{2\over M_j} + {1\over M_j} \sum_{j=2}^{M_j-1} P_{Y|X}(D_i|x_i)\label{eq:li_3}\\
		 &\le& {2\over M_j} + {1\over M_j} \sum_{j=2}^{M_j-1} \left( 1-
2Q\left(\mathrm{Leb}(D_i)\over 2\right) \right)\label{eq:li_4}\\
		&\le& {2\over M_j} + {M_j - 2\over M_j} \left(1 -  
2Q\left({1\over 2M_j-4}\sum_{i=2}^{M_j-1}\mathrm{Leb}(D_i)\right) \right)\label{eq:li_5}\\
		&\le& {2\over M_j} + {M_j - 2\over M_j} \left(1 -  
2Q\left({\sqrt{nP}\over M_j-2}\right) \right)\label{eq:li_6}\\
		&\le& {2\over M_j} + {\sqrt{2nP\over \pi}\over M_j}\label{eq:li_7}\,,
\end{eqnarray}
where in~\eqref{eq:li_2} $P_{Y|X=x} = \matn(x, 1)$,~\eqref{eq:li_3} follows by
upper-bounding probability of successful decoding for $i=1$ and $i=M_j$ by 1,~\eqref{eq:li_4}
follows since clearly for a fixed value of the length  $\mathrm{Leb}(D_i)$ the optimal location of
the interval $D_i$, maximizing the value $P_{Y|X}(D_i|x_i)$, is centered at $x_i$,~\eqref{eq:li_5} is by Jensen's inequality applied to $x\to
1-2Q(x)$ concave for $x\ge 0$,~\eqref{eq:li_6} is because
\begin{equation}
	\bigcup_{i=2}^{M_j-1} D_i \subset [-\sqrt{nP}, \sqrt{nP}]
\end{equation}
and $D_i$ are disjoint, and~\eqref{eq:li_7} is by
\begin{equation}
	1 - 2Q(x) \le \sqrt{2\over \pi} x\,,\qquad x\ge0.
\end{equation}
Thus,~\eqref{eq:li_7} completes the proof of~\eqref{eq:li_f},~\eqref{eq:li_0} and the theorem.
\end{IEEEproof}

\begin{IEEEproof}[Proof of Theorem~\ref{th:lp_new1}]
Notice that for any $0\le \lambda \le P$ we have
\begin{equation}\label{eq:lup1_1}
	 C(P-\lambda) \le C(P) - {\lambda \log e\over 2(1+P)}\,. 
\end{equation}
On the other hand, by concavity of $\sqrt{V(P)}$ and since $V(0)=0$ we have for any $0\le \lambda
\le P$
\begin{equation}\label{eq:lpn1_2}
	 \sqrt{V(P-\lambda)} \ge \sqrt{V(P)} - {\sqrt{V(P)}\over P}\lambda\,.
\end{equation}
Thus, taking $s=\lambda n$ in Lemma~\ref{th:linf} we get with the help of~\eqref{eq:lup1_1} and~\eqref{eq:lpn1_2}:
\begin{equation}\label{eq:lpn1_3}
	 \PP[\norm{\vect x}^2_\infty \ge s] \le 
		\exp\left\{ \Delta_n  - (b_1 - b_2
n^{-{1\over 2}}) s\right\}\,,
\end{equation}
where we denoted for convenience
\begin{eqnarray} b_1 &=& {\log e \over 2 (1+P)}\,,\\
   b_2 &=& {\sqrt{V(P)}\over P} Q^{-1}(\epsilon)\,,\\
 \Delta_n &=& nC(P) - \sqrt{nV(P)} Q^{-1}(\epsilon) + 2\log n -  \log M + \log b\,. 
\end{eqnarray}
 Note that Lemma~\ref{th:linf} only shows validity of~\eqref{eq:lpn1_3} for $0 \le s \le nP$, but since
for $s>nP$ the left-hand side is zero, the statement actually holds for all $s\ge 0$. Then for $n\ge
N(P,\epsilon)$ we have
\begin{equation}
	(b_1 - b_2 n^{-{1\over 2}})\ge {b_1 \over 2}
\end{equation}
and thus further upper-bounding~\eqref{eq:lpn1_3} we get
\begin{equation}\label{eq:lpn1_4}
	 \PP[\norm{\vect x}^2_\infty \ge s] \le 
		\exp\left\{ \Delta_n - {b_1 s\over
2}\right\}\,.
\end{equation}
Finally, if the code is so large that $\Delta_n < 0$,
then~\eqref{eq:lpn1_4} would imply that $\PP[\norm{\vect x}^2_\infty \ge s] <1$ for all $s\ge 0$, 
which is clearly impossible. Thus we must have $\Delta_n \ge 0$ for any $(n, M, \epsilon)_{max,det}$
code. The proof concludes by taking $s = {2(\log 2 + \Delta_n)\over b_1}$ in~\eqref{eq:lpn1_4}.
\end{IEEEproof}

\begin{IEEEproof}[Proof of Theorem~\ref{th:lp_new2}] To prove~\eqref{eq:lpn2} we will show the
following statement: {\it There exist two constants $b_0$ and $b_1$ such that for any $(n, M_1, \epsilon)$ code for the $AWGN(P)$ channel with codewords
$\vect x$ satisfying
\begin{equation}\label{eq:lu3_0}
	 \norm{\vect x}_4 \ge b n^{1\over 4} 
\end{equation}
we have an upper bound on the cardinality:
\begin{equation}\label{eq:lu3_1}
	 M_1 \le {4\over 1-\epsilon} \exp\left\{nC + 2 \sqrt{ nV\over 1-\epsilon} -
b_1 (b-b_0)^2\sqrt{n}\right\}\,,
\end{equation}
provided $b \ge b_0(P, \epsilon)$}. From here~\eqref{eq:lpn2} follows by first upper-bounding
$(b-b_0)^2 \ge {b^2\over 2} - b_0^2$ and then verifying easily that the choice 
\begin{equation}
	b^2 = {2\over b_1\sqrt{n}}(nC + b_2 \sqrt{n} - \log {M\over 2})
\end{equation}
with $b_2 = b_0^2 b_1 + 2\sqrt{V\over 1-\epsilon} + \log {4\over 1-\epsilon}$
takes the right-hand side of~\eqref{eq:lu3_1} below $\log {M\over 2}$.

To prove~\eqref{eq:lu3_1},  denote
\begin{equation}
	S = b - \left(6\over 1+\epsilon\right)^{1\over 4}
\end{equation}
and choose $b$ large enough so that
\begin{equation}\label{eq:lu3_5a}
	 \delta \eqdef S - 6^{1\over 4} \sqrt{1+P} > 0\,.
\end{equation}
Then, on one hand we have
\begin{eqnarray} 
P_{Y^n}[ \norm{Y^n}_4 \ge S n^{1\over 4}] &=& 
	\PP[ \norm{X^n + Z^n}_4 \ge S n^{1\over 4}]\\
	&\ge& \PP[ \norm{X^n}_4 - \norm{Z^n}_4 \ge S n^{1\over 4}]\label{eq:lu3_2}\\
	&\ge& \PP[ \norm{Z^n}_4 \le n^{1\over 4}(S-b)]\label{eq:lu3_3}\\
	&\ge& {1+\epsilon\over 2} \label{eq:lu3_4}\,,
\end{eqnarray}
where~\eqref{eq:lu3_2} is by the triangle inequality for $\norm{\cdot}_4$,~\eqref{eq:lu3_3} is by the
constraint~\eqref{eq:lu3_0} and~\eqref{eq:lu3_4} is by the Chebyshev inequality applied to
$\norm{Z^n}_4^4 = \sum_{j=1}^n Z_j^4$.  On the other hand, we have
\begin{eqnarray} 
	P_{Y^n}^*[ \norm{Y^n}_4 \le S n^{1\over 4}] &=&
	P_{Y^n}^*[ \norm{Y^n}_4 \le (6^{1\over 4} \sqrt{1+P} + \delta) n^{1\over 4}]
		\label{eq:lu3_5}\\
	&\ge& P_{Y^n}^*[ \{\norm{Y^n}_4 \le 6^{1\over 4} \sqrt{1+P} n^{1\over 4}\} +
		\{\norm{Y^n}_4 \le \delta n^{1\over 4}\}]\label{eq:lu3_6}\\
	&\ge& P_{Y^n}^*[ \{\norm{Y^n}_4 \le 6^{1\over 4} \sqrt{1+P} n^{1\over 4}\} +
		\{\norm{Y^n}_2 \le \delta n^{1\over 4}\}]\label{eq:lu3_7}\\
	&\ge& 1 - \exp\{-b_1 \delta^2 \sqrt{n}\}\label{eq:lu3_8}\,,
\end{eqnarray}
where~\eqref{eq:lu3_5} is by the definition of $\delta$ in~\eqref{eq:lu3_5a},~\eqref{eq:lu3_6} is by
the triangle inequality for $\norm{\cdot}_4$ which implies the inclusion
\begin{equation}
	\{\vect y\colon \norm{\vect y}_4 \le a+b\} \supset \{\vect y\colon \norm{\vect y}_4 \le a\} +
\{\vect y\colon \norm{\vect y}_4 \le b\}
\end{equation}
with $+$ denoting the Minkowski sum of sets,~\eqref{eq:lu3_7} is by~\eqref{eq:al_qhigh2} with $p=2$,
$q=4$; and~\eqref{eq:lu3_8} holds for some $b_1 = b_1(P)>0$ by the Gaussian isoperimetric
inequality~\cite{ST74} which is applicable since
\begin{equation}
	P_{Y^n}^*[\norm{Y^n}_4 \le 6^{1\over 4} \sqrt{1+P} n^{1\over 4}] \ge {1\over 2}
\end{equation}
by the Chebyshev inequality applied to $\sum_{j=1}^n Y_j^4$ (note: $Y^n\sim \matn(0,
1+P)^n$ under $P_{Y^n}^*$). As a side remark, we add that the estimate of the 
large-deviations of the sum of $4$-th powers of iid Gaussians as $\exp\{-O(\sqrt{n})\}$ is order-optimal.

Together~\eqref{eq:lu3_4} and~\eqref{eq:lu3_8} imply
\begin{equation}\label{eq:lu3_11}
	 \beta_{1+\epsilon\over 2}(P_{Y^n}, P_{Y^n}^*) \le \exp\{-b_1 \delta^2 \sqrt{n}\}\,. 
\end{equation}

On the other hand, by~\cite[Lemma 59]{PPV08} we have for any $\vect x$ with $\norm{\vect x}_2\le
\sqrt{nP}$ and any $0<\alpha<1$:
\begin{equation}\label{eq:lu3_9}
	 \beta_\alpha(P_{Y^n | X^n=\vect x}, P_{Y^n}^*) \ge {\alpha\over 2} \exp\left\{-nC -
\sqrt{2nV\over \alpha}\right\}\,,
\end{equation}
where $C$ and $V$ are the capacity and the dispersion of the $AWGN(P)$ channel. Then, by convexity
in $\alpha$ of the right-hand side of~\eqref{eq:lu3_9} and~\cite[Lemma 32]{YP10} we have for any
input distribution $P_{X^n}$:
\begin{equation}\label{eq:lu3_10}
	 \beta_\alpha(P_{X^n Y^n}, P_{X^n} P_{Y^n}^*) \ge {\alpha\over 2} \exp\left\{-nC -
\sqrt{2nV\over \alpha}\right\}\,.
\end{equation}

We complete the proof of~\eqref{eq:lu3_1} by invoking Theorem~\ref{th:ss_meta} in the
form~\eqref{eq:ssavg} with $Q_Y=P_{Y^n}^*$ and $\alpha = {1+\epsilon\over 2}$:
\begin{equation}\label{eq:lu3_12}
	 \beta_{1+\epsilon\over2}(P_{Y^n}, P_{Y^n}^*) \ge 
		M_1 \beta_{1 - \epsilon\over 2}(P_{X^nY^n}, P_{X^n} P_{Y^n}^*)\,.
\end{equation}
Applying bounds~\eqref{eq:lu3_11} and~\eqref{eq:lu3_10} to~\eqref{eq:lu3_12}  we conclude
that~\eqref{eq:lu3_1} holds with
\begin{equation}
	b_0 = \left(6\over 1+\epsilon\right)^{1\over 4}+6^{1\over 4} \sqrt{1+P}\,.
\end{equation}

Next, we proceed to the proof of~\eqref{eq:awgn4}. On one hand, we have
\begin{eqnarray} \sum_{j=1}^n\EE\left[ Y_j^4\right] &=&
		\sum_{j=1}^n \EE\left[(X_j + Z_j)^4\right]\label{eq:lu3_13}\\
		&=& \sum_{j=1}^n \EE[ X_j^4 + 6 X_j^2 Z_j^2 + Z_j^4]\label{eq:lu3_14}\\
		&\le& \EE[\norm{\vect x}_4^4] + 6nP + 3n\,,\label{eq:lu3_15}
\end{eqnarray}
where~\eqref{eq:lu3_13} is by the definition of the AWGN channel,~\eqref{eq:lu3_14} is because $X^n$
and $Z^n$ are independent and thus odd terms vanish,~\eqref{eq:lu3_15} is by the power-constraint
$\sum X_j^2 \le nP$. On the other hand, applying Proposition~\ref{th:dvc_iid} with $f(y) = - y^4$,
$\theta=2$ and using~\eqref{eq:re1} we obtain\footnote{Of course, a similar Gaussian lower bound
holds for any cumulative sum, in particular for any power $\sum \EE[|Y_j|^q], q \ge 1$.}
\begin{equation}\label{eq:lu3_16}
	\sum_{j=1}^n\EE\left[ Y_j^4\right] \ge 3n(1+P)^2 - (nC  - \log M + b_3 \sqrt{n}) n^{1\over 4}\,,
\end{equation}
for some $b_3 = b_3(P, \epsilon)>0$. Comparing~\eqref{eq:lu3_16} and~\eqref{eq:lu3_15}
statement~\eqref{eq:awgn4} follows.

We remark that by extending Proposition~\ref{th:dvc_iid} to expectations like ${1\over n-1}
\sum_{j=1}^{n-1}\EE[Y_j^2 Y_{j+1}^2]$, cf.~\eqref{eq:dvi_remark}, we could provide a lower bound
similar to~\eqref{eq:awgn4} for more general 4-th degree polynomials in $\vect x$. For example, 
it is possible to treat the case of $p(\vect x) = \sum_{i,j} a_{i-j} x_i^2 x_j^2$, 
where $A=(a_{i-j})$ is a Toeplitz positive semi-definite matrix. We would proceed as
in~\eqref{eq:lu3_15}, computing $\EE[p(Y^n)]$ in two ways, with the only difference that 
the peeled off quadratic polynomial would require application of Theorem~\ref{th:qft} 
instead of the simple power constraint.
Finally, we also mention that the method~\eqref{eq:lu3_15} does not work for estimating
$\EE[\norm{\vect x}_6^6]$ because we would need an \textit{upper} bound $\EE[\norm{\vect
x}_4^4]\simleq 3nP^2$, which is not possible to obtain in the context of $O(\sqrt{n})$-achieving
codes as the counterexamples~\eqref{eq:lp_low1} show.

\end{IEEEproof}
\fi 

\bibliographystyle{IEEEtran}
\bibliography{IEEEabrv,reports}
\end{document}